\documentclass[10pt]{article}
\usepackage[utf8]{inputenc}
\usepackage{geometry}
\usepackage{titling}
\usepackage{authblk}
\usepackage{physics}
\usepackage{amssymb}
\usepackage{amsmath}
\usepackage{amsthm}
\allowdisplaybreaks[4]
\usepackage{slashed}
\usepackage{multirow}
\usepackage{graphicx,color,epsfig}
\usepackage{cite}
\usepackage[colorlinks=true,linkcolor=red,citecolor=blue]{hyperref}
\usepackage{placeins}

\newcommand{\wm}{\phantom{-}}

\begin{document}
\title{Investigation of $\Lambda_c \to (\Lambda,n)\ell^+ \nu_\ell $ Decays in Standard Model and Beyond}
\author{Zhi-Hua Zhang, Ying Li$^1$ \footnote{liying@ytu.edu.cn}}
\affil{Department of Physics, Yantai University, Yantai 264005, China}
\maketitle
\vspace{0.2cm}

\begin{abstract}
In this work, we study the decays $\Lambda_c \to (\Lambda, n) \ell^+ \nu_\ell$ ($\ell = \mu, e$) within a model-independent framework. We calculate the helicity amplitudes for all possible four-fermion operators, including the interactions between different new physics (NP) operators. The form factors for the $\Lambda_c \to (\Lambda, n)$ transitions are taken from lattice QCD calculations. We present detailed results for the branching fractions and other key observables. Although our results are generally consistent with previous studies, we find that the predicted central values for the branching fractions are approximately $10\%$ larger than the experimental measurements. Additionally, we explore the potential impacts of NP, focusing particularly on the scenario in which NP particles couple exclusively to the muon. Using Wilson coefficients fitted from $D$ and $D_s$ meson decays, we examine NP effects in the $\Lambda_c \to (\Lambda, n) \mu^+ \nu_\mu$ decay channels. It is found that although most potential contributions of NP are obscured by the uncertainties inherent in both theory and experiment, the right-handed scalar operator could suppress the branching fraction of $\Lambda_c \to (\Lambda, n) \mu^+ \nu_\mu$ by up to $10\%$. We also highlight that the ratio of the forward-backward asymmetry in the $\Lambda_c \to (\Lambda, n) \mu^+ \nu_\mu$ decay to that in the $\Lambda_c \to (\Lambda, n) e^+ \nu_e$ decay provides a novel probe for NP, as it is less sensitive to hadronic uncertainties and is largely unaffected by current NP operators. All of our results can be tested in ongoing experiments such as BESIII, Belle-II, and LHCb, as well as in future high-energy facilities like the Super Tau-Charm Factory (STCF) and the Circular Electron Positron Collider (CEPC).
\end{abstract}

\section{Introduction} \label{sec:1}
The study of charm physics has become a critical area of focus in modern particle physics due to the charm quark's unique position at the intersection of weak and strong interactions. As a quark heavier than up, down, and strange quarks, yet lighter than bottom and top quarks, the charm quark serves as an essential probe for exploring both perturbative and non-perturbative aspects of QCD. One of the key strengths of the charm sector is its ability to provide precise tests of the Standard Model (SM), while simultaneously offering fertile ground for investigating the possible existence of new physics (NP) beyond the SM. The study of charm quarks also enables the accurate measurement of fundamental parameters such as the Cabibbo-Kobayashi-Maskawa (CKM) matrix elements, which govern quark mixing and CP violation in weak decays \cite{Richman:1995wm,Bianco:2003vb,Cerri:2018ypt, Li:2021iwf}. The weak decays of charm mesons, such as the $D$ -mesons, and charm baryons, such as the $\Lambda_c$, provide an excellent opportunity to test the predictions of SM regarding quark mixing and CP violation with high precision \cite{Achasov:2023gey, Cheng:2021qpd}. These decays, particularly those that result in hadronic final states, offer stringent tests of factorization models within QCD, thereby refining our understanding of both perturbative and non-perturbative QCD effects. Moreover, charm physics provides a complementary framework to the study of heavier quarks, such as the bottom and top quarks. The charm sector offers experimental advantages in terms of precision and accessibility, making it a valuable alternative for conducting high-precision measurements that are often more challenging in the bottom and top sectors. Beyond its utility in testing the SM, the charm sector holds significant promise for the search for new physics. Deviations from SM predictions in charm decays could signal the presence of new physics, including phenomena such as lepton-flavor violating processes, or exotic decays. The precision with which charm decays can be measured, combined with the relatively clean experimental environment provided by charmed meson and baryon decays, enables rigorous constraints to be placed on a wide array of NP models. Ongoing and upcoming experiments, including those at BESIII, Belle-II, LHCb, and other high-luminosity facilities, are expected to dramatically improve the sensitivity of these searches. These advancements promise not only to deepen our understanding of charm physics but also to provide critical insights into the potential discovery or exclusion of NP, solidifying the pivotal role of the charm sector as an essential pillar of modern particle physics research.

In recent years, the charged-current decays of $\overline B \to D^{(*)} \ell^- \bar\nu_\ell$ have been extensively studied by the BaBar, Belle, and LHCb collaborations \cite{ParticleDataGroup:2024cfk}, and they have produced results that challenge the SM predictions. A particular focus has been placed on the ratios $R(D^{(*)})$, defined as the ratio of the branching fractions:
\begin{eqnarray}
R(D^{(*)}) \equiv \dfrac{{\cal B}(\overline B \to D^{(*)} \tau^- \bar\nu_\tau)}{{\cal B}(\overline B \to D^{(*)}\ell^- \bar\nu_\ell)},
\end{eqnarray}
where $\ell=e, \mu$. The experimental results show that these ratios exceed their SM predictions by more than three standard deviations ($\sigma$) \cite{HFLAV:2022esi}. This large deviation from the SM is suggestive of a potential violation of lepton flavor universality (LFU), a principle that posits that all leptons, including electrons, muons, and tau, should interact identically in weak decays. This discrepancy has sparked a wave of interest and investigation within the particle physics community, with many theorists and experimentalists exploring whether the observed tension could be attributed to NP beyond the SM \cite{Li:2018lxi, London:2021lfn, Capdevila:2023yhq, Iguro:2024hyk,Bhutta:2025cia}.

The potential violation of LFU observed in $B$-meson decays has naturally prompted the extension of the search for similar NP signatures to the charm sector. Specifically, the leptonic and semi-leptonic decays of charm mesons, such as $D$ and $D_s$, induced by the weak transitions $c\to (s,d) \ell^+ \nu_\ell$, provide valuable probes for both SM predictions and possible NP effects. Numerous theoretical studies have explored these decays within the SM as well as in various NP scenarios \cite{Wang:2014uiz, Fajfer:2015ixa, Soni:2018adu, Fleischer:2019wlx, Leng:2020fei, Faustov:2019mqr}. These studies predict the expected behavior of these decays in the presence of NP interactions, and comparing these predictions with experimental results will provide crucial insights into potential deviations from the SM. Theoretical predictions in this area are eagerly awaited to be tested in current and upcoming experiments, notably those at BES-III, LHCb, and Belle-II. In addition to the decays of mesons, another promising approach to studying $c\to (s,d) \ell^+ \nu_\ell$ transitions is through the semi-leptonic decays of charm baryons. These processes offer an alternative and complementary perspective, as they involve the weak decay of baryons rather than mesons, providing a distinct probe for flavor physics. Typical examples of such decays include $\Lambda_c \to \Lambda \ell^+ \nu_\ell$ \cite{Alvarado:2024lpq} and $\Lambda_c \to n \ell^+ \nu_\ell$ \cite{Meinel:2016dqj}. These semi-leptonic decays serve as additional tests of LFU, while also opening new experimental opportunities for the investigation of NP in the charm sector. Furthermore, studying charm-baryon decays is crucial for precise determination of the CKM matrix elements $|V_{cs}|$ or $|V_{cd}|$ \cite{Bolognani:2024cmr}. By combining experimental measurements of decay rates for $\Lambda_c \to (\Lambda,n) \ell^+ \nu_\ell$ with lattice QCD calculations of the corresponding form factors, these decays provide an excellent opportunity to extract these CKM elements with high precision. Moreover, from the perspective of an experimentalist, $\Lambda_c \to (\Lambda,n) \ell^+ \nu_\ell$  decay modes are used as normalization modes in measurements of a wide range of other charm and bottom baryon decays \cite{Rosner:2012gj}.

In recent years, significant progress has been made in measuring the semileptonic decays of the $ \Lambda_c $ baryon ~\cite{BESIII:2015ysy, BESIII:2016ffj, BESIII:2022ysa, BESIII:2023jxv}, specifically the modes $ \Lambda_c\to\Lambda \ell \nu_\ell $. These efforts, primarily by the BESIII collaboration, have yielded increasingly precise results. The most accurate branching fractions to date are $ \mathcal{B}\left(\Lambda_c \to \Lambda e^{+} \nu_e\right)=(3.56 \pm 0.11 \pm 0.07 ) \% $ ~\cite{BESIII:2022ysa} and $ \mathcal{B}\left(\Lambda_c \to \Lambda \mu^{+} \nu_\mu\right)=(3.48 \pm 0.14 \pm 0.10 ) \% $~\cite{BESIII:2023jxv}. Comparing the electron mode with the inclusive semileptonic decay rate $ \mathcal{B}\left(\Lambda_c \to X e^{+} \nu_e\right)=(3.95 \pm 0.34 \pm 0.09 ) \% $, it is apparent that additional exclusive decay channels remain to be measured. In fact, BESIII has reported evidence for decays involving excited $\Lambda$ states, $ \Lambda_c \to \Lambda(1520) e^{+} \nu_e $ and $ \Lambda_c \to \Lambda(1405) e^{+} \nu_e $~\cite{BESIII:2022qaf}, although with small branching fractions of $(1.02 \pm 0.52 \pm 0.11 ) \times 10^{-3} $ and $ (0.42 \pm 0.19 \pm 0.04 ) \times 10^{-3} $, respectively. While a range of decays has been explored, experimental data for the $ \Lambda_c\to n \ell \nu_\ell $ semileptonic modes, also predicted by the SM, is still lacking. Theoretical investigations of the $\Lambda_c \to (\Lambda, n) \ell^+ \nu_\ell$ decays have been extensively carried out using various approaches \cite{Faustov:2019ddj, Gutsche:2015rrt, Huang:2021ots, Li:2016qai, Azizi:2009wn, Khodjamirian:2011jp, Li:2021qod, Zhao:2018zcb, Geng:2020gjh, Gutsche:2014zna, Pervin:2005ve, Faustov:2016yza, Lu:2016ogy, Geng:2019bfz, He:2021qnc, Geng:2020fng, Meinel:2016dqj, Meinel:2017ggx, Bahtiyar:2021voz, Zhang:2023nxl}. A comparison between the current experimental results and theoretical predictions reveals that while most SM predictions align with the experimental data, there remain significant uncertainties on both sides. As a result, the possibility of NP effects cannot be completely ruled out at this stage.

In our previous study \cite{Leng:2020fei}, we had examined the decays of $D$ mesons by assuming general effective Hamiltonians that describe the $c \to (s,d) \ell^+ \nu_\ell$ transitions, incorporating the full set of four-fermion operators. Using the most recent experimental data, we performed a minimum $\chi^2$-fit of the Wilson coefficients corresponding to each operator and calculated various observables for the pure-leptonic and semi-leptonic decays of $D$ mesons. In this work, we shall extend our approach to investigate the semi-leptonic decays $\Lambda_c^+ \to \Lambda \ell^+ \nu_\ell$ and $\Lambda_c^+ \to n \ell^+ \nu_\ell$ within the same framework. We will analyze these decays both in the context of the SM and with respect to possible contributions from NP. This paper will provide a comprehensive theoretical treatment of these decays, including the relevant form factors and decay dynamics. Additionally, we will review current experimental results and discuss their implications for both SM physics and NP scenarios, with particular focus on potential signals of NP and LFU violation.

This work is organized as follows. In Sec. \ref{sec:II} , we present the framework of our study, including the effective Hamiltonian, form factors, and helicity amplitudes. The SM predictions and NP analysis  are listed in Sec.~\ref{sec:III}. Finally, we summarize the main findings and conclude in Sec.~\ref{sec:4}.
\section{Analysis Formula} \label{sec:II}
\subsection{Effective Hamiltonian of $c\to (s,d) \ell^+\nu_\ell$ Transitions}
Consider the decay process $\Lambda_c\to \Lambda \ell^+\nu_\ell$ as an example. Assuming the absence of right-handed neutrinos, the effective Hamiltonian for the $c\to s \ell^+\nu_\ell$ transition can be written as \cite{Fajfer:2015ixa}:
\begin{eqnarray}\label{eH}
 {\cal H}_{eff}=\frac{G_{F}}{\sqrt{2}}V_{cs}\Big[(1+C_{VL})O_{VL}+C_{VR}O_{VR}+C_{SL}O_{SL}+C_{SR}O_{SR}+C_{T}O_{T}\Big]+h.c.
\end{eqnarray}
Here, $G_{F}$ is the Fermi coupling constant, $V_{cs}$ is the CKM matrix element, and $C_{i}(i=VL,VR,SL,SR,T)$ represent the Wilson coefficients. The $O_{i}$ are four-fermion operators with different chiral and Lorentz structures. The explicit forms of the operators $O_{i}$, categorized by their left- and right-handed chiralities, are provided as follows:
\begin{eqnarray}
   &O_{VL}=(\bar{s}\gamma^{\mu}P_{L}c)(\bar{\nu}_\ell\gamma_{\mu}P_{L}\ell),\,\,\,\,
   O_{VR}=(\bar{s}\gamma^{\mu}P_{R}c)(\bar{\nu}_{\ell}\gamma_{\mu}P_{L}\ell),\,\,\,\,\nonumber\\
    &O_{SL}=(\bar{s}P_{L}c)(\bar{\nu}_{\ell}P_{R}\ell),\,\,\,\,
    O_{SR}=(\bar{s}P_{R}c)(\bar{\nu}_{\ell}P_{R}\ell) ,\,\,\,\,\nonumber\\
    &O_{T}=(\bar{s}\sigma^{\mu\nu}P_{R}c)(\bar{\nu}_{\ell}\sigma_{\mu\nu}P_R\ell),
\end{eqnarray}
with $P_{L,R}=1\mp\gamma_5$. It should be noted that the operator $O_{VL}$ is present only in the SM and the Wilson coefficients of $O_{X}$ can get modified at the scale of short distance. We also note that the tensor operator with opposite quark chiralities vanishes within the appropriate Fierz identity.

Therefore, the transition of the process of $\Lambda_c\to \Lambda \ell^{+}\nu_\ell$ can be written as
\begin{eqnarray}\label{m}
    \cal{M}&=&\langle \Lambda\ell^{+}\nu_{\ell}|{\cal H}_{eff}|\Lambda_c\rangle
    \nonumber\\
    &=&\frac{G_F}{\sqrt{2}}V_{cs}\Bigg\{\Big[(1+C_{VL})\bra{\Lambda}\bar{s}\gamma^{\mu}P_Lc\ket{\Lambda_c}+C_{VR}\bra{\Lambda}\bar{s}\gamma^{\mu}P_Rc\ket{\Lambda_c}\Big]\langle\ell^{+}\nu_\ell|\bar{\nu}_{\ell}\gamma_{\mu}P_L\ell|0\rangle\nonumber\\ &&+\Big[C_{SL}\bra{\Lambda}\bar{s}P_Lc\ket{\Lambda_c}+C_{SR}\bra{\Lambda}\bar{s}P_Rc\ket{\Lambda_c}\Big]\langle\ell^{+}\nu_\ell|\bar{\nu}_{\ell}P_R\ell|0\rangle\nonumber\\    &&+C_{T}\bra{\Lambda}\bar{s}\sigma^{\mu\nu}P_Rc\ket{\Lambda_c}\langle\ell^{+}\nu_\ell|\bar{\nu}_{\ell}\sigma_{\mu\nu}P_R\ell|0\rangle\Bigg\}.
\end{eqnarray}

\subsection{Form factor of hadron matrix element}
To calculate the hadron helicity amplitudes, it is necessary for us to parameterize the hadron matrix element of $\Lambda_c\to\Lambda$. The transition  $\Lambda_c\to\Lambda$  involving vector and axial-vector currents can be expressed in terms of six form factors, which are given by \cite{Feldmann:2011xf,Datta:2017aue}
\begin{eqnarray}
   \langle \Lambda(p_2,\lambda_2)|\bar{s}\gamma^{\mu}c|\Lambda_c(p_1,\lambda_1)\rangle &=&\bar{u}_2(p_2,\lambda_2)\Bigg[f_0(q^2)(M_{\Lambda_c}-M_\Lambda)\frac{q^\mu}{q^2} \nonumber\\
   &&+f_+(q^2)\frac{M_{\Lambda_c}+M_\Lambda}{Q_+}\bigg(p_1^{\mu}
    +p_2^{\mu}-(M_{\Lambda_c}^2-M_\Lambda^2)\frac{q^{\mu}}{q^2}\bigg)\nonumber \\
   &&+f_{\perp}(q^2)\bigg(\gamma^{\mu}-\frac{2M_{\Lambda}}{Q_+}p^{\mu}_1-\frac{2M_{\Lambda_c}}{Q_+}p_2^{\mu}\bigg)\Bigg]u_1(p_1,\lambda_1),\\
  \langle \Lambda(p_2,\lambda_2)|\bar{s}\gamma^{\mu}\gamma_5c|\Lambda_c(p_1,\lambda_1)\rangle
   &=&-\bar{u}_2(p_2,\lambda_2)\gamma_5\Bigg[g_0(q^2)(M_{\Lambda_c}+M_\Lambda)\frac{q^{\mu}}{q^2}\nonumber\\
 &&+g_+(q^2)\frac{M_{\Lambda_c}-M_\Lambda}{Q_-}\bigg(p_1^{\mu}+p_2^{\mu}
    -(M_{\Lambda_c}^2-M_\Lambda^2)\frac{q^{\mu}}{q^2}\bigg)\nonumber\\
  &&+g_{\perp}(q^2)\bigg(\gamma^{\mu}+\frac{2M_\Lambda}{Q_-}p_1^{\mu}-\frac{2M_{\Lambda_c}}{Q_-}p_2^{\mu}\bigg)\Bigg]u_1(p_1,\lambda_1),
\end{eqnarray}
with $q=p_1-p_2$. $\lambda_i=\pm\frac{1}{2}\,(i=1,2)$ indicate the helicities of initial hadron and the final one, respectively. Within the equation of motion, the hadronic matrix elements involving the scalar and pseudoscalar currents can be parameterized as
\begin{eqnarray}
   \langle \Lambda(p_2,\lambda_2)|\bar{s}c|\Lambda_c(p_1,\lambda_1)\rangle&=&f_0(q^2)
   \frac{M_{\Lambda_c}-M_\Lambda}{m_c-m_d}\bar{u}_2(p_2,\lambda_2)u_1(p_1,\lambda_1),
   \label{eq:scalar ffs}\\
    \langle \Lambda(p_2,\lambda_2)|\bar{s}\gamma_5 c|\Lambda_c(p_1,\lambda_1)\rangle &=& g_0(q^2) \frac{M_{\Lambda_c}+M_\Lambda}{m_c+m_d} \bar{u}_2(p_2,\lambda_2)\gamma_5u_1(p_1,\lambda_1),\label{eq:pseudo-scalar ffs}
\end{eqnarray}
and $M_\pm=M_{\Lambda_c}\pm M_\Lambda$. The matrix elements of the tensor currents can be written in terms of four form factors $h_+$, $h_\perp$, $\widetilde{h}_+$, $\widetilde{h}_\perp$ \cite{Feldmann:2011xf,Datta:2017aue},
\begin{eqnarray}
&&\langle \Lambda(p_2,\lambda_2)| \bar{s}i\sigma^{\mu\nu} c|\Lambda_c(p_1,\lambda_1)\rangle
=\bar{u}_2(p_2,\lambda_2)\Big[2h_+(q^2)\frac{p_{1}^\mu p_{2}^{ \nu}-p_{1}^\nu p_{2}^{\mu}}{Q_+} \nonumber\\
&&+h_\perp (q^2)\Big(\frac{M_{\Lambda_c}+M_{\Lambda}}{q^2}(q^\mu \gamma^\nu -q^\nu \gamma^\mu)
-2(\frac{1}{q^2}+\frac{1}{Q_+})(p_{1}^\mu p_{2}^{\nu}-p_{1}^\nu p_{2}^{\mu}) \Big) \nonumber\\
&&+\widetilde{h}_+ (q^2)\Big(i\sigma^{\mu \nu}-\frac{2}{Q_-}(M_{\Lambda_c}(p_{2}^{\mu}\gamma^\nu -p_{2}^{\nu}\gamma^\mu)-M_{\Lambda}(p_{1}^\mu \gamma^\nu -p_{1}^\nu \gamma^\mu)+p_{1}^\mu p_{2}^{\nu}-p_{1}^\nu p_{2}^{\mu}) \Big) \nonumber\\
&&+\widetilde{h}_\perp(q^2) \frac{M_{\Lambda_c}-M_{\Lambda}}{q^2 Q_-}\Big((M_{\Lambda_c}^2-M_{\Lambda}^2-q^2)(\gamma^\mu p_{1}^\nu - \gamma^\nu p_{1}^\mu)-(M_{\Lambda_c}^2-M_{\Lambda}^2+q^2)(\gamma^\mu p_{2}^{\nu}-\gamma^\nu p_{2}^{\mu})\nonumber\\
&&+2(M_{\Lambda_c}-M_{\Lambda})(p_{1}^\mu p_{2}^{\nu}-p_{1}^\nu p_{2}^{\mu}) \Big)
\Big]u_1(p_1,\lambda_1), \label{eq:TFF}
\end{eqnarray}
with $Q_\pm =(M_{\Lambda_c} \pm M_{\Lambda})^2 - q^2$.
The matrix elements of the current $\bar{c}i\sigma^{\mu\nu}\gamma_5b$ can be obtained from the above equation by using the identity
\begin{eqnarray}
\sigma^{\mu \nu}\gamma_{5}=-\frac{i}{2}\epsilon^{\mu \nu \alpha \beta}\sigma_{\alpha \beta}.
\end{eqnarray}
We note that the above parameterizations decompose the matrix elements into form factors within the helicity basis. In the literature, these matrix elements are also often parameterized using the so-called ``Weinberg form factors" \cite{Weinberg:1958ut, Gutsche:2015mxa, Chen:2001zc}. The relationship between those form factors and the ones used in the present context can be found in Ref.~\cite{Mu:2019bin}.

For these introduced helicity form factors, they are always expanded by \cite{Meinel:2016dqj,Meinel:2017ggx}
\begin{eqnarray}
f(q^2) &=& \frac{1}{1-q^2/(m_{\rm pole}^f)^2} \sum_{n=0}^{\rm {max}}a_n^f\big[z(q^2)\big]^n, \label{eq:nominalfitphys}
\end{eqnarray}
where the function $z(q^2)$ is given as
\begin{equation}
z(q^2) = \frac{\sqrt{t_+ -q^2}-\sqrt{t_+ -t_0}}{\sqrt{t_+ -q^2}+\sqrt{t_+ -t_0}},
\end{equation}
with $t_0=(M_{\Lambda_c}-M_{\Lambda})^2$, $t^f_+=(m_D+m_{K})^2$. The quantum numbers and masses of the $D$ mesons producing these poles in the different form factors are given in Table.~\ref{tab:polemasses}. The fit results for the parameters of the $\Lambda_c \to \Lambda$ and $\Lambda_c \to n$ transitions are results base on the lattice QCD approach, which are summarized in Table.~\ref{tab:fitresults1} and \ref{tab:fitresults2}, respectively. The nominal fit is used to evaluate the central values and statistical uncertainties, while the higher-order fit is used
to estimate systematic uncertainties.

\begin{table}[!htb]
\centering
\caption{\label{tab:polemasses}
 The quantum numbers and masses of the $D$ mesons producing poles in the different form factors \cite{Meinel:2016dqj,Meinel:2017ggx}.}
\begin{tabular}{c|ccccc}
\hline\hline
 Transition & $f$     & \hspace{1ex} & $J^P$ & \hspace{1ex}  & $m_{\rm pole}^f$ [GeV]  \\
 \hline
 \multirow{4}{*}{$\Lambda_c \to \Lambda$}
&$f_+$, $f_\perp$                         && $1^-$   && $2.112$  \\
&$f_0$                                                      && $0^+$   && $2.318$  \\
&$g_+$, $g_\perp$ && $1^+$   && $2.460$  \\
&$g_0$                                                      && $0^-$   && $1.968$  \\
\hline
 \multirow{4}{*}{$\Lambda_c \to n$}
&$f_+$, $f_\perp$, $h_+$, $h_\perp$                         && $1^-$   && $2.010$  \\
&$f_0$                                                      && $0^+$   && $2.351$  \\
&$g_+$, $g_\perp$, $\widetilde{h}_+$, $\widetilde{h}_\perp$ && $1^+$   && $2.423$  \\
&$g_0$                                                      && $0^-$   && $1.870$  \\
\hline\hline
\end{tabular}
\end{table}

\begin{table}
\centering
\caption{Results for the $z$-expansion parameters describing the form factors of $\Lambda_c \to \Lambda$ \cite{Meinel:2016dqj}.} \label{tab:fitresults1}
\begin{tabular}{ccccc|ccccc}
\hline\hline
   $a_i^f$ & &  Nominal fit  &  & Higher-order fit& $a_i^f$ & &  Nominal fit  &  & Higher-order fit \\
\hline
   $a_0^{f_\perp}$     && $\wm 1.30\pm 0.06$ && $\wm 1.28\pm 0.07$&
$a_1^{f_\perp}$     &&    $-3.27\pm 1.18$ &&    $-2.85\pm 1.34$ \\
$a_2^{f_\perp}$     && $\wm 7.16\pm 11.6$ && $\wm 7.14\pm 12.2$ &
$a_3^{f_\perp}$     &&                    &&    $-1.08\pm 30.0$ \\
\hline
$a_0^{f_+}$         && $\wm 0.81\pm 0.03$ && $\wm 0.79\pm 0.04$ &
$a_1^{f_+}$         &&    $-2.89\pm 0.52$ &&    $-2.38\pm 0.61$ \\
$a_2^{f_+}$         && $\wm 7.82\pm 4.53$ && $\wm 6.64\pm 6.07$ &
$a_3^{f_+}$         &&                    &&    $-1.08\pm 29.8$ \\
\hline
$a_0^{f_0}$         && $\wm 0.77\pm 0.02$ && $\wm 0.76\pm 0.03$ &
$a_1^{f_0}$         &&    $-2.24\pm 0.51$ &&    $-1.77\pm 0.58$ \\
$a_2^{f_0}$         && $\wm 5.38\pm 4.80$ && $\wm 4.93\pm 6.28$ &
$a_3^{f_0}$         &&                    &&    $-0.26\pm 29.8$ \\
\hline
$a_0^{g_\perp}$ && $\wm 0.68\pm 0.02$ && $\wm 0.67\pm 0.02$ &
$a_1^{g_\perp}$     &&    $-1.91\pm 0.35$ &&    $-1.73\pm 0.54$ \\
$a_2^{g_\perp}$     && $\wm 6.24\pm 4.89$ && $\wm 5.97\pm 6.64$ &
$a_3^{g_\perp}$     &&                    &&    $-1.68\pm 29.8$ \\
\hline
$a_0^{g_+}$ && $\wm 0.68\pm 0.02$ && $\wm 0.67\pm 0.02$ &
$a_1^{g_+}$         &&    $-2.44\pm 0.25$ &&    $-2.22\pm 0.35$ \\
$a_2^{g_+}$         && $\wm 13.7\pm 2.15$ && $\wm 12.1\pm 4.43$ &
$a_3^{g_+}$         &&                    && $\wm 12.9\pm 29.2$ \\
\hline
$a_0^{g_0}$         && $\wm 0.71\pm 0.03$ && $\wm 0.72\pm 0.04$ &
$a_1^{g_0}$         &&    $-2.86\pm 0.44$ &&    $-2.80\pm 0.53$ \\
$a_2^{g_0}$         && $\wm 11.8\pm 2.47$ && $\wm 11.7\pm 4.74$ &
$a_3^{g_0}$         &&                    && $\wm 1.35\pm 29.4$ \\
\hline
\hline
\end{tabular}
\end{table}

\begin{table}[!htb]
\centering
\caption{Results for the $z$-expansion parameters describing the form factors of $\Lambda_c \to n$ \cite{Meinel:2017ggx}.}\label{tab:fitresults2}
\begin{tabular}{ccccc|ccccc}
\hline\hline
 $a_i^f$ & &  Nominal fit  &  & Higher-order fit&
 $a_i^f$ & &  Nominal fit  &  & Higher-order fit \\
\hline
$a_0^{f_\perp}$ & & $\wm 1.36\pm 0.07$ & & $\wm 1.32\pm 0.09$&
$a_1^{f_\perp}$ & & $-1.70\pm 0.83$ & & $-1.33\pm 0.98$ \\
$a_2^{f_\perp}$ & & $\wm 0.71\pm 4.34$ & & $-1.38\pm 8.60$ &
$a_3^{f_\perp}$ & &  & & $\wm 7.02\pm 29.2$ \\
\hline
$a_0^{f_+}$ & & $\wm 0.83\pm 0.04$ & & $\wm 0.80\pm 0.05$ &
$a_1^{f_+}$ & & $-2.33\pm 0.56$ & & $-1.94\pm 0.83$ \\
$a_2^{f_+}$ & & $\wm 8.41\pm 3.05$ & & $\wm 5.33\pm 8.04$ &
$a_3^{f_+}$ & &  & & $\wm 10.1\pm 28.8$ \\
\hline
$a_0^{f_0}$ & & $\wm 0.84\pm 0.04$ & & $\wm 0.82\pm 0.05$ &
$a_1^{f_0}$ & & $-2.57\pm 0.60$ & & $-2.42\pm 0.88$ \\
$a_2^{f_0}$ & & $\wm 9.87\pm 3.15$ & & $\wm 7.71\pm 8.19$ &
$a_3^{f_0}$ & &  & & $\wm 9.30\pm 28.8$ \\
\hline
$a_0^{g_\perp}$ & & $\wm 0.69\pm 0.02$ & & $\wm 0.68\pm 0.03$ &
$a_1^{g_\perp}$ & & $-0.68\pm 0.32$ & & $-0.89\pm 0.58$ \\
$a_2^{g_\perp}$ & & $\wm 0.70\pm 2.18$ & & $\wm 3.97\pm 6.81$ &
$a_3^{g_\perp}$ & &  & & $-10.8\pm 25.2$ \\
\hline
$a_0^{g_+}$ & & $\wm 0.69\pm 0.02$ & & $\wm 0.68\pm 0.03$ &
$a_1^{g_+}$ & & $-0.90\pm 0.29$ & & $-1.07\pm 0.55$ \\
$a_2^{g_+}$ & & $\wm 2.25\pm 1.90$ & & $\wm 3.46\pm 6.42$ &
$a_3^{g_+}$ & &  & & $\wm 0.49\pm 24.1$  \\
\hline
$a_0^{g_0}$ & & $\wm 0.73\pm 0.04$ & & $\wm 0.71\pm 0.05$ &
$a_1^{g_0}$ & & $-0.97\pm 0.52$ & & $-0.93\pm 0.77$ \\
$a_2^{g_0}$ & & $\wm 0.83\pm 2.61$ & & $\wm 1.64\pm 7.87$ &
$a_3^{g_0}$ & &  & & $-1.73\pm 28.0$\\
\hline
$a_0^{h_\perp}$ & & $\wm 0.63\pm 0.03$ & & $\wm 0.62\pm 0.05$ &
$a_1^{h_\perp}$ & & $-1.04\pm 0.45$ & & $-0.88\pm 0.72$ \\
$a_2^{h_\perp}$ & & $\wm 1.42\pm 2.67$ & & $\wm 1.42\pm 7.74$&
$a_3^{h_\perp}$ & &  & & $-0.41\pm 27.8$  \\
\hline
$a_0^{h_+}$ & & $\wm 1.11\pm 0.07$ & & $\wm 1.10\pm 0.10$&
$a_1^{h_+}$ & & $-0.69\pm 0.92$ & & $-0.56\pm 1.07$ \\
$a_2^{h_+}$ & & $-2.84\pm 5.19$ & & $-3.85\pm 9.28$ &
$a_3^{h_+}$ & &  & & $\wm 5.61\pm 29.5$ \\
\hline
$a_0^{\widetilde{h}_\perp}$ & & $\wm 0.63\pm 0.03$ & & $\wm 0.63\pm 0.05$ &
$a_1^{\widetilde{h}_\perp}$ & & $-1.39\pm 0.58$ & & $-1.55\pm 0.81$ \\
$a_2^{\widetilde{h}_\perp}$ & & $\wm 4.22\pm 3.97$ & & $\wm 6.20\pm 8.12$ &
$a_3^{\widetilde{h}_\perp}$ & &  & & $-5.19\pm 28.3$ \\
\hline
$a_0^{\widetilde{h}_+}$ & & $\wm 0.63\pm 0.03$ & & $\wm 0.63\pm 0.05$ &
$a_1^{\widetilde{h}_+}$ & & $-1.19\pm 0.56$ & & $-1.23\pm 0.80$ \\
$a_2^{\widetilde{h}_+}$ & & $\wm 3.73\pm 3.73$ & & $\wm 4.36\pm 8.08$ &
$a_3^{\widetilde{h}_+}$ & &  & & $-0.84\pm 28.1$ \\
\hline\hline
\end{tabular}

\end{table}

\subsection{Decay Amplitudes}
Now, we turn to calculate the decay rate of decay  $\Lambda_c \to \Lambda\ell^+ \nu_\ell $, which is relate to $|{\cal M}|^2$. In SM, the transitions $c\to s \ell^{+}\nu_l$ can be viewed as subsequent processes $c\to s W^{*+}$ and $W^{*+}\to \ell^{+}\nu_l$ subsequently. It is known to us that the off-shell $W^{*+}$ has four helicities, namely $\lambda_W=\pm 1,0\,(J=1)$ and $\lambda_W=0\,(J=0)$, and the off-shell $W^{*+}$ has a time-like polarization, with $J=1,0$ denoting the two angular momenta of the rest frame $W^{*+}$. In order to distinguish the two $\lambda_{W^{*}}=0$ states we adopt the notation $\lambda_{W^{*}}=0$ for $J=1$ and $\lambda_{W^{*}}=t$ for $J=0$. In the $\Lambda_c$-baryon rest frame, choosing the $z$--axis to be along the $W^{*+}$, the polarization vectors of virtual particle are:
\begin{align} \label{epsilon}
  \varepsilon^{\mu}(t)=(1,0,0,0) , \quad \varepsilon^{\mu}(\pm)=\frac{1}{\sqrt{2}}(0,\mp,-i,0),\quad \varepsilon^{\mu}(0)=(0,0,0,1),
\end{align}
where $q^\mu$ is the four-momentum of the off-shell $W^{*+}$. In this case, the polarization vectors of the $W^{*+}$ satisfy the orthonormality and completeness relation
\begin{align}\label{eq:completeness}
    \sum_{\lambda_{W_1},\lambda_{W_2}}\epsilon^{\dag}_{\mu}(\lambda_{W_1})
    \epsilon_{\mu^\prime}(\lambda_{W_2})g_{\lambda_{W_1}\lambda_{W_2}}=g_{\mu\mu^\prime},
\end{align}
with $g_{\lambda_{W_1}\lambda_{W_2}}={\rm diag}(+,-,-,-)$ for $\lambda_{W_1},\lambda_{W_2}=t,\pm,0$.

\begin{figure}[htb]
\begin{center}
\includegraphics[width=0.5\textwidth]{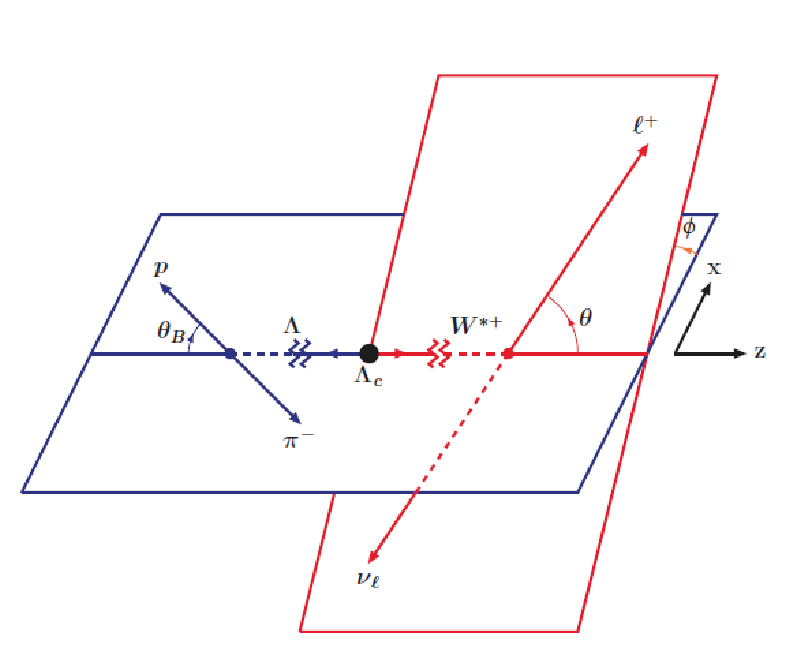}
\caption{Definition of the polar and the azimuthal angles.}
\label{fig:angles}
\end{center}
\end{figure}

It is much convenient to calculate the helicity amplitudes in the rest frame of the parent baryon $\Lambda_c$, where we choose the $z$-axis to be along the $W^{*+}$ (see Fig.~\ref{fig:angles}). In the experimental side, the polarization of $\Lambda$ can be probed by analyzing the angular decay distribution of the subsequent decay of $\Lambda$. As an exemplary case, we shall consider the decay mode $\Lambda \to p\pi^-$ as a polarization analyzer. One can exploit the cascade nature of the decay $\Lambda_c \to \Lambda (\to p\pi^-) W^{*+}(\to \ell^{+}\nu_l) $ by writing down a joint angular decay distribution involving the polar angles $\theta$, $\theta_B$ and the azimuthal angles $\phi$ defined by the decay products in their respective center of mass systems as shown in Figure.~\ref{fig:angles}.

The hadronic amplitudes that represent the processes $\Lambda_c \to \Lambda W^{*+}$ with vector and axial-vector currents are defined by
\begin{eqnarray}
&&{H}_{\lambda_2,\lambda_W}^V=\varepsilon^*_\mu(\lambda_W)\langle \Lambda(p_2,\lambda_2)|\bar{s}\gamma^{\mu}c |\Lambda_c(p_1,\lambda_1)\rangle, \\
&&{H}_{\lambda_2,\lambda_W}^A= \varepsilon^*_\mu(\lambda_W)\langle \Lambda(p_2,\lambda_2)|\bar{s}\gamma^{\mu}\gamma_5 c |\Lambda_c(p_1,\lambda_1)\rangle;
\end{eqnarray}
and we then have
\begin{eqnarray}
{H}_{\lambda_2,\lambda_W}^{L(R)}={\cal H}_{\lambda_2,\lambda_W}^V\mp{\cal H}_{\lambda_2,\lambda_W}^A
= \varepsilon^*_\mu(\lambda_W) \langle \Lambda(p_2,\lambda_2) |\bar{s}\gamma^{\mu}(1\mp\gamma_5)c|\Lambda_c(p_1,\lambda_1)\rangle.
\end{eqnarray}
Due to the conservation of the angular momentum, one has $\lambda_W=\lambda_1+\lambda_2$. With the convention of eq. (\ref{epsilon}) and the definitions of form factors, we thus obtain the hadron helicity amplitudes as
\begin{align} \label{eq:hel_inv}
    &H^L_{\frac{1}{2},t}=f_0(q^2)M_-\sqrt{\frac{Q_+}{q^2}}-g_0(q^2)M_+\sqrt{\frac{Q_-}{q^2}}, \\
    &H^L_{\frac{1}{2},+}=f_\perp(q^2)\sqrt{2Q_-}-g_\perp(q^2)\sqrt{2Q_+}, \\
    &H^L_{\frac{1}{2},0}=f_+(q^2)M_+\sqrt{\frac{Q_-}{q^2}}-g_+(q^2)M_-\sqrt{\frac{Q_+}{q^2}},\\
    &H^L_{-\frac{1}{2},t}=f_0(q^2)M_-\sqrt{\frac{Q_+}{q^2}}+g_0(q^2)M_+\sqrt{\frac{Q_-}{q^2}},\\
    &H^L_{-\frac{1}{2},-}=f_\perp(q^2)\sqrt{2Q_-}+g_\perp(q^2)\sqrt{2Q_+}, \\
    &H^L_{-\frac{1}{2},0}=f_+(q^2)M_+\sqrt{\frac{Q_-}{q^2}}+g_+(q^2)M_-\sqrt{\frac{Q_+}{q^2}}, \\
    &H^R_{\frac{1}{2},t}=f_0(q^2)M_-\sqrt{\frac{Q_+}{q^2}}+g_0(q^2)M_+\sqrt{\frac{Q_-}{q^2}}, \\
    &H^R_{\frac{1}{2},+}=f_\perp(q^2)\sqrt{2Q_-}+g_\perp(q^2)\sqrt{2Q_+}, \\
    &H^R_{\frac{1}{2},0}=f_+(q^2)M_+\sqrt{\frac{Q_-}{q^2}}+g_+(q^2)M_-\sqrt{\frac{Q_+}{q^2}}, \\
    &H^R_{-\frac{1}{2},t}=f_0(q^2)M_-\sqrt{\frac{Q_+}{q^2}}-g_0(q^2)M_+\sqrt{\frac{Q_-}{q^2}}, \\
    &H^R_{-\frac{1}{2},-}=f_\perp(q^2)\sqrt{2Q_-}-g_\perp(q^2)\sqrt{2Q_+}, \\
    &H^R_{-\frac{1}{2},0}=f_+(q^2)M_+\sqrt{\frac{Q_-}{q^2}}-g_+(q^2)M_-\sqrt{\frac{Q_+}{q^2}}.
\end{align}

We also define the hadron helicity amplitudes with scalar currents as
\begin{eqnarray}
{H}^{SPL(R)}_{\lambda_2}= \langle{\Lambda(p_2,\lambda_2)}|\bar{s}(1\mp\gamma_5)c|\Lambda_c(p_1,\lambda_1)\rangle,
\end{eqnarray}
and the non-zero helicity amplitudes are given as
\begin{align}\label{eq:scalar1 amplitudes}
    &H^{SPL}_{\frac{1}{2}}=f_0(q^2)\frac{M_-}{m_c-m_s}\sqrt{Q_+}+g_0(q^2)\frac{M_+}{m_c+m_s}\sqrt{Q_-}, \\
    &H^{SPL}_{-\frac{1}{2}}=f_0(q^2)\frac{M_-}{m_c-m_s}\sqrt{Q_+}-g_0(q^2)\frac{M_+}{m_c+m_s}\sqrt{Q_-}, \\
    &H^{SPR}_{\frac{1}{2}}=f_0(q^2)\frac{M_-}{m_c-m_s}\sqrt{Q_+}-g_0(q^2)\frac{M_+}{m_c+m_s}\sqrt{Q_-}, \\
    &H^{SPR}_{-\frac{1}{2}}=f_0(q^2)\frac{M_-}{m_c-m_s}\sqrt{Q_+}+g_0(q^2)\frac{M_+}{m_c+m_s}\sqrt{Q_-}.
\end{align}
Similarly, the hadron helicity amplitude with tensor operator is defined as
\begin{eqnarray}
{H}^{T}_{\lambda_2,\lambda_{W_1},\lambda_{W_2}}
=\varepsilon^{*}_\mu(\lambda_{W_1})\varepsilon^{*}_\nu(\lambda_{W_2})
\langle \Lambda(p_2,\lambda_2) |\bar{s}i\sigma^{\mu\nu}(1+\gamma_5)c|\Lambda_c(p_1,\lambda_1)\rangle,
\end{eqnarray}
with the relation $\lambda_{W_1}+\lambda_{W_2}=\lambda_1+\lambda_2$. The non-zero amplitudes are given as
\begin{align}\label{eq:tensor amplitudes}
     &H^T_{-\frac{1}{2},t,0}=h_{+}(q^2)\sqrt{Q_{-}}+\Tilde{h}_{+}(q^2)\sqrt{Q_{+}}, \\
     &H^T_{\frac{1}{2},t,0}=h_{+}(q^2)\sqrt{Q_{+}}-\Tilde{h}_{+}(q^2)\sqrt{Q_{+}}, \\
     &H^T_{\frac{1}{2},t,+1}=-h_{\perp}(q^2)M_{+}\sqrt{\frac{2Q_{-}}{q^2}}+\Tilde{h}_{\perp}(q^2)M_{-}\sqrt{\frac{2Q_{+}}{q^2}}, \\
     &H^T_{-\frac{1}{2},t,-1}=-h_{\perp}(q^2)M_{+}\sqrt{\frac{2 Q_{-}}{q^2}} -\Tilde{h}_{\perp}(q^2)M_{-}\sqrt{\frac{2 Q_{+}}{q^2}},\\
     &H^T_{\frac{1}{2},0,+1}=-h_{\perp}(q^2)M_{+}\sqrt{\frac{2Q_{-}}{q^2}}+\Tilde{h}_{\perp}(q^2)M_{-}\sqrt{\frac{2Q_{+}}{q^2}},\\
     &H^T_{-\frac{1}{2},0,-1}=h_{\perp}(q^2)M_{+}\sqrt{\frac{2Q_{-}}{q^2}}+\Tilde{h}_{\perp}(q^2)M_{-}\sqrt{\frac{2Q_{+}}{q^2}},\\
     &H^T_{\frac{1}{2},+,-}=-h_{+}(q^2)\sqrt{Q_{-}}-\Tilde{h}_{+}\sqrt{Q_{+}},\\
     &H^T_{-\frac{1}{2},+,-}=-h_{+}(q^2)\sqrt{Q_{-}}+\Tilde{h}_{+}\sqrt{Q_{+}},
\end{align}
and the relationship $H^T_{\lambda_2,\lambda_{W_1},\lambda_{W_2}}=-H^T_{\lambda_2,\lambda_{W_2},\lambda_{W_1}}$ holds due to the parity conservation.

For the lepton part $W^{*+}\to \ell^{+}\nu_l$, the helicity amplitudes are:
\begin{align}
&L^{L}_{\lambda_\ell,\lambda_W}=\varepsilon^{\mu}(\lambda_W)\bar{\nu}_{\ell}(-\frac{1}{2})\gamma_{\mu}(1-\gamma_5)\ell(\lambda_\ell),\\
    &L^{SPR}_{\lambda_\ell}=\bar{\nu}_{\ell}(-\frac{1}{2})(1+\gamma_5)\ell(\lambda_\ell),\\
    &L^T_{\lambda_\ell,\lambda_{W_1},\lambda_{W_2}}=-i\varepsilon^{\mu}(\lambda_{W_1})
    \varepsilon^{\nu}(\lambda_{W_2})\bar{\nu}_{\ell}(-\frac{1}{2})\sigma_{\mu\nu}(1+\gamma_5)\ell(\lambda_\ell),
\end{align}
where $\lambda_\ell$ stands for the helicity of the lepton. After calculation, the non-vanishing lepton helicity amplitudes are also given as
\begin{eqnarray} \label{eq:lel_inv}
  & &L^{SPR}_{-\frac{1}{2}}=2e^{i\phi}\beta\sqrt{q^2}, \\
  & &L^{L}_{-\frac{1}{2},t}=-2e^{i\phi}\beta m_\ell, \\
  & &L^{L}_{-\frac{1}{2},0}=-2e^{i\phi}\beta m_\ell \cos{\theta},\\
  & &L^{L}_{-\frac{1}{2},+}=\sqrt{2}e^{2i\phi}\beta m_\ell\sin{\theta},\\
  & &L^{L}_{-\frac{1}{2},-}=-\sqrt{2}\beta m_\ell\sin{\theta}, \\
  & &L^{L}_{\frac{1}{2},0}=-2\beta\sqrt{q^2} \sin{\theta}, \\
  & &L^{L}_{\frac{1}{2},+}=-e^{i\phi}\beta \sqrt{2q^2}(1+\cos{\theta}),\\
  & &L^{L}_{\frac{1}{2},-}=-e^{-i\phi}\beta \sqrt{2q^2}(1-\cos{\theta}), \\
  & &L^T_{\frac{1}{2},0,+}=-L^T_{\frac{1}{2},+,t}=-\sqrt{2}e^{i\phi}\beta m_{\ell}(1+\cos{\theta}),\\
  & &L^T_{\frac{1}{2},0,-}= L^T_{\frac{1}{2},-,t}=\sqrt{2}e^{-i\phi}\beta m_{\ell} (1-\cos{\theta}), \\
  & &L^T_{\frac{1}{2},t,0}=-L^T_{\frac{1}{2},+,-}=-2\beta m_{\ell}  \sin{\theta},\\
  & &L^T_{-\frac{1}{2},0,+}=e^{2i\phi}\beta\sqrt{2q^2} \sin{\theta},\\
  & &L^T_{-\frac{1}{2},0,-}=\beta\sqrt{2q^2}  \sin{\theta},  \\
  & &L^T_{-\frac{1}{2},+,t}=-e^{2i\phi}\beta\sqrt{2q^2} \sin{\theta},\\
  & &L^T_{-\frac{1}{2},-,t}= \beta\sqrt{2q^2}\sin{\theta}, \\
  & &L^T_{-\frac{1}{2},t,0}=-L^T_{-\frac{1}{2},+,-}=-2e^{i\phi}\beta\sqrt{q^2}\cos{\theta},
\end{eqnarray}
and $L^T_{\lambda_\ell,\lambda_{W_1},\lambda_{W_2}}=-L^T_{\lambda_\ell,\lambda_{W_2},\lambda_{W_1}}$, with
\begin{eqnarray}
\beta=\sqrt{1-\frac{m_\ell^2}{q^2}}.
\end{eqnarray}

From Eq.(\ref{m}), we the have
\begin{eqnarray}
\sum_{\lambda_1,\lambda_2,\lambda_\ell}|\mathcal{M}|^2&=&\frac{G_F^2|V_{cs}|^2}{2}\sum_{\lambda_2,\lambda_\ell}
\bigg[|1+C_{VL}|^2|\mathcal{T}_{VL}|^2+|C_{VR}|^2|\mathcal{T}_{VL}|^2
 +|C_{SL}|^2|\mathcal{T}_{SL}|^2+|C_{SR}|^2|\mathcal{T}_{SR}|^2\nonumber\\
  &  &+|C_T|^2|\mathcal{T}_T|^2+(1+C_{VL})C_{VR}^{\dag}\mathcal{T}_{VL}\mathcal{T}_{VR}^\dag+(1+C_{VL})^{\dag}C_{VR}\mathcal{T}_{VR}\mathcal{T}_{VL}^\dag
  \nonumber\\
  & &+(1+C_{VL})C_{SL}^{\dag}\mathcal{T}_{VL}\mathcal{T}_{SL}^{\dag}+(1+C_{VL})^{\dag}C_{SL}\mathcal{T}_{SL}\mathcal{T}_{VL}^{\dag}
  +(1+C_{VL})C_{SR}^{\dag}\mathcal{T}_{VL}\mathcal{T}_{SR}^{\dag}\nonumber\\
  &  &+(1+C_{VL})^{\dag}C_{SR}\mathcal{T}_{SR}\mathcal{T}_{VL}^{\dag}+(1+C_{VL})C_T^{\dag}\mathcal{T}_{VL}\mathcal{T}_T^{\dag}
  +(1+C_{VL})^{\dag}C_T\mathcal{T}_T\mathcal{T}_{VL}^{\dag}\nonumber\\
  &  &+C_{VR}C_{SL}^{\dag}\mathcal{T}_{VR}\mathcal{T}_{SL}^{\dag}+C_{VR}^{\dag}C_{SL}\mathcal{T}_{SL}\mathcal{T}_{VR}^{\dag}
 +C_{VR}C_{SR}^{\dag}\mathcal{T}_{VR}\mathcal{T}_{SR}^{\dag}+C_{VR}^{\dag}C_{SR}\mathcal{T}_{SR}\mathcal{T}_{VR}^{\dag}\nonumber\\
  &  &+C_{VR}C_T^{\dag}\mathcal{T}_{VR}\mathcal{T}_T^{\dag}+C_TC_{VR}^{\dag}\mathcal{T}_T\mathcal{T}_{VR}^{\dag}
      +C_{SL}C_{SR}^{\dag}\mathcal{T}_{SL}\mathcal{T}_{SR}^{\dag}+C_{SR}C_{SL}^{\dag}\mathcal{T}_{SR}\mathcal{T}_{SL}^{\dag}\nonumber\\
  &  &+C_{SL}C_T^{\dag}\mathcal{T}_{SL}\mathcal{T}_T^{\dag}+C_TC_{SL}^{\dag}\mathcal{T}_T\mathcal{T}_{SL}^{\dag}
      +C_{SR}C_T^{\dag}\mathcal{T}_{SR}\mathcal{T}_T^{\dag}+C_TC_{SR}^{\dag}\mathcal{T}_T\mathcal{T}_{SR}^{\dag}\bigg],
\end{eqnarray}
where $\mathcal{T}_i=\langle \Lambda\ell^{+}\nu_{\ell}|O_i|\Lambda_c\rangle$, $i=VL,VR,SL,SR,T$. For the first term in the bracket, we use the completeness relation~(\ref{eq:completeness}) and calculate $|\mathcal{T}_{VL}|^2$ as following,
\begin{eqnarray}
  \sum_{\lambda_i}|\mathcal{T}_{VL}|^2
   &=&\sum_{\lambda_i}\langle \Lambda\ell^{+}\nu_{\ell}|O_{VL}|\Lambda_c\rangle \langle \Lambda\ell^{+}\nu_{\ell}| O_{VL}|\Lambda_c\rangle^{\dag}\nonumber\\
   &=&\sum_{\lambda_i} \langle \Lambda|\bar s\gamma^\mu P_L c|\Lambda_c\rangle \langle \ell^{+}\nu_{\ell}|\bar{\nu}_\ell\gamma_{\mu}P_{L}\ell|0\rangle \langle \Lambda|\bar s\gamma^\nu P_L c|\Lambda_c\rangle^\dag \langle \ell^{+}\nu_{\ell}|\bar{\nu}_\ell\gamma_{\nu}P_{L}\ell|0\rangle^\dag \nonumber\\
   &=&\sum_{\lambda_i} \langle \Lambda|\bar s\gamma^\mu P_L c|\Lambda_c\rangle \langle \Lambda|\bar s\gamma^\nu P_L c|\Lambda_c\rangle^\dag
     \langle \ell^{+}\nu_{\ell}|\bar{\nu}_\ell\gamma^{\mu^\prime}P_{L}\ell|0\rangle  \langle \ell^{+}\nu_{\ell}|\bar{\nu}_\ell\gamma^{\nu^\prime}P_{L}\ell|0\rangle^\dag g_{\mu\mu^\prime}g_{\nu\nu^\prime}
    \nonumber\\
   &=&\sum_{\lambda_i} \sum_{\lambda_{W_i}}\langle \Lambda|\bar s\gamma^\mu P_L c|\Lambda_c\rangle \langle \Lambda|\bar s\gamma^\nu P_L c|\Lambda_c\rangle^\dag
     \langle \ell^{+}\nu_{\ell}|\bar{\nu}_\ell\gamma^{\mu^\prime}P_{L}\ell|0\rangle  \langle \ell^{+}\nu_{\ell}|\bar{\nu}_\ell\gamma^{\nu^\prime}P_{L}\ell|0\rangle^\dag  \nonumber\\
   &   &\varepsilon_{\mu}^{\dag}(\lambda_{W_1})\varepsilon_{\mu^\prime}(\lambda_{W_2})g_{\lambda_{W_1}\lambda_{W_2}}\varepsilon_{\nu}(\lambda_{W_3})
   \varepsilon^{\dag}_{\nu^\prime}(\lambda_{W_4})g_{\lambda_{W_3}\lambda_{W_4}}
    \nonumber\\
   & = &\sum_{\lambda_i} \sum_{\lambda_{W_i}}\langle \Lambda|\bar s\gamma^\mu P_L c|\Lambda_c\rangle \varepsilon_{\mu}^{\dag}(\lambda_{W_1})\langle \Lambda|\bar s\gamma^\nu P_L c|\Lambda_c\rangle^\dag \varepsilon_{\nu}(\lambda_{W_3})\nonumber\\
   &  & \langle \ell^{+}\nu_{\ell}|\bar{\nu}_\ell\gamma^{\mu^\prime}P_{L}\ell|0\rangle \varepsilon_{\mu^\prime}(\lambda_{W_2}) \langle \ell^{+}\nu_{\ell}|\bar{\nu}_\ell\gamma^{\nu^\prime}P_{L}\ell|0\rangle^\dag   \varepsilon^{\dag}_{\nu^\prime}(\lambda_{W_4})
    g_{\lambda_{W_1}\lambda_{W_2}}g_{\lambda_{W_3}\lambda_{W_4}}
    \nonumber\\
 & =&\sum_{\lambda_i}\sum_{\lambda_{W_i}}H^L_{\lambda_2,\lambda_{W_1}}
 (H^L_{\lambda_2,\lambda_{W_3}})^{\dag}L^{L}_{\lambda_\ell,\lambda_{W_2}}
(L^{L}_{\lambda_\ell,\lambda_{W_4}})^{\dag}g_{\lambda_{W_1}\lambda_{W_2}}g_{\lambda_{W_3} \lambda_{W_4}}.
\end{eqnarray}
The four factors in the last line are Lorentz-invariant and can therefore be evaluated in different Lorentz frames. The leptonic part is evaluated in the center-of-mass (CM) frame of the lepton and neutrino pair $(\ell,\nu_\ell)$ or equivalently, the rest frame of the off-shell $W_{\rm off-shell}$, with the positive $z$-axis aligned along the direction of the off-shell $W$, introducing the decay angle $\theta$, as shown in Eqs.(\ref{eq:lel_inv}). The hadronic part is evaluated in the $\Lambda_b$ rest frame bringing in the helicity amplitudes, which are also shown in Eqs.~(\ref{eq:hel_inv}).

\subsection{Observables}

In above calculation,  we sum over the helicities of both the parent and daughter baryon. The three-body decay can be described in terms of the invariant variable $q^2$ and the polar angle $\theta$, as defined in Figure.~\ref{fig:angles}. The differential $(q^2,\cos\theta)$ distribution reads
\begin{eqnarray}
\frac{d^2\Gamma(\Lambda_c\to \Lambda \ell^{+}\nu_\ell)}{dq^2 d\cos\theta} =  \frac{G_F^2|V_{cs}|^2q^2\sqrt{Q_+Q_-}}{2^{10}\pi^3M_{\Lambda_c}^3}
\Big(1-\frac{m_
\ell^2}{q^2}\Big)^2 {\cal A}_{\rm total},\label{eq:distr2}
\end{eqnarray}
with
\begin{eqnarray} \label{eq:tamp}
 {\cal A}_{\rm total}
 &=&|1+C_{VL}|^2{\cal A}_{VL}
     +|C_{VR}|^2{\cal A}_{VR}
     +|C_{SL}|^2{\cal A}_{SL}
     +|C_{SR}|^2{\cal A}_{SR}
      +|C_{T}|^2{\cal A}_{T}\nonumber\\
 & &+2{\rm Re}[(1+C_{VL})^*C_{VR}]{\cal A}^{\rm int}_{VL,VR}
    +2{\rm Re}[(1+C_{VL})^*C_{SL}]{\cal A}^{\rm int}_{VL,SL}
    +2{\rm Re}[(1+C_{VL})^*C_{SR}]{\cal A}^{\rm int}_{VL,SR}\nonumber\\
 & &+2{\rm Re}[(1+C_{VL})^*C_{T}]{\cal A}^{\rm int}_{VL,T}
    +2{\rm Re}[C_{VR}^*C_{SL}]{\cal A}^{\rm int}_{VR,SL}
    +2{\rm Re}[C_{VR}^*C_{SR}]{\cal A}^{\rm int}_{VR,SR}
\nonumber\\
 & &+2{\rm Re}[C_{VR}^*C_{T}]{\cal A}^{\rm int}_{VR,T}
    +2{\rm Re}[C_{SL}^*C_{SR}]{\cal A}^{\rm int}_{SL,SR}
    +2{\rm Re}[C_{SL}^*C_{T}]{\cal A}^{\rm int}_{SL,T}
    +2{\rm Re}[C_{SR}^*C_{T}]{\cal A}^{\rm int}_{SR,T}.
\end{eqnarray}
The functions are given as:
\begin{eqnarray}
 {\cal A}_{VL}
    &=& 2\sin^2{\theta} \left(|H^L_{\frac{1}{2},0}|^2+|H^L_{-\frac{1}{2},0}|^2 \right )
        +(1+\cos{\theta})^2|H^L_{\frac{1}{2},+}|^2+(1-\cos{\theta})^2|H^L_{-\frac{1}{2},-}|^2 \nonumber\\
    & & +\frac{{m_\ell}^2}{q^2} \Bigg[2\cos^2{\theta}\left(|H^L_{\frac{1}{2},0}|^2+|H^L_{-\frac{1}{2},0}|^2\right )
        +\sin^2{\theta}\left(| H^L_{\frac{1}{2},+}|^2+|H^L_{-\frac{1}{2},-}|^2 \right)\nonumber\\
    & & +2\left(|H^L_{\frac{1}{2},t} |^2 +|H^L_{-\frac{1}{2},t}|^2 \right)
        -4\cos{\theta}\left(H^L_{\frac{1}{2},t}H^L_{\frac{1}{2},0}+ H^L_{-\frac{1}{2},t} H^L_{-\frac{1}{2},0}\right)\Bigg ];\\
 {\cal A}_{VR}
    &=&2 \sin^2{\theta} \left(|H^R_{\frac{1}{2},0}|^2 +| H^R_{-\frac{1}{2},0}|^2 \right )
        +(1+\cos{\theta})^2 |H^R_{\frac{1}{2},+}|^2+(1-\cos{\theta})^2|H^R_{-\frac{1}{2},-}|^2 \nonumber\\
    & & +\frac{{m_\ell}^2}{q^2}\Bigg[2\cos^2{\theta}\left(|H^R_{\frac{1}{2},0}|^2+|H^R_{-\frac{1}{2},0}|^2\right )
        +\sin^2{\theta}\left(|H^R_{\frac{1}{2},+}|^2+|H^R_{-\frac{1}{2},-}|^2 \right)\nonumber\\
    & & +2\left(|H^R_{\frac{1}{2},t}|^2+|H^R_{-\frac{1}{2},t}|^2\right)
        -4\cos{\theta}\left(H^R_{\frac{1}{2},t}H^R_{\frac{1}{2},0}+H^R_{-\frac{1}{2},t}H^R_{-\frac{1}{2},0}\right)\Bigg];\\
 {\cal A}_{SL}
    &=&2 \left(|H^{SPL}_{-\frac{1}{2}}|^2+|H^{SPL}_{\frac{1}{2}}|^2\right);\\
 {\cal A}_{SR}
    &=&2 \left(|H^{SPR}_{-\frac{1}{2}}|^2+|H^{SPR}_{\frac{1}{2}}|^2\right);\\
 {\cal A}_{T}
    &=& 4\sin^2{\theta}\Big[(H^T_{\frac{1}{2},+,t}-H^T_{\frac{1}{2},+,0})^2+(H^T_{-\frac{1}{2},0,-}-H^T_{-\frac{1}{2},-,t})^2\Big]\nonumber\\
    & & +8\cos^2{\theta}\Big[(H^T_{-\frac{1}{2},+,-}-H^T_{-\frac{1}{2},0,t})^2+(H^T_{\frac{1}{2},+,-}-H^T_{\frac{1}{2},0,t})^2\Big]\nonumber\\
    & & +\frac{4m_\ell^2}{q^2}\Bigg\{(1-\cos{\theta})^2\left(H^T_{-\frac{1}{2},0,-}-H^T_{-\frac{1}{2},-,t}\right)^2
        +(1+\cos{\theta})^2\left(H^T_{\frac{1}{2},+,t}-H^T_{\frac{1}{2},+,0}\right)^2\nonumber\\
    & & +2\sin^2{\theta}\Big[\left(H^T_{-\frac{1}{2},+,-}-H^T_{-\frac{1}{2},0,t}\right)^2
        + \left(H^T_{\frac{1}{2},+,-}-H^T_{\frac{1}{2},0,t}\right)^2\Big]\Bigg\};\\
 {\cal A}_{VL,VR}^{\rm int}
   &=&2\sin^2{\theta}\Big(H^L_{-\frac{1}{2},0}H^R_{-\frac{1}{2},0}+H^L_{\frac{1}{2},0}H^R_{\frac{1}{2},0}\Big)
       +(1+\cos{\theta})^2H^L_{\frac{1}{2},+}H^R_{\frac{1}{2},+}+(1-\cos{\theta})^2H^L_{-\frac{1}{2},-}H^R_{-\frac{1}{2},-}\nonumber\\
   & & +\frac{m_\ell^2}{q^2}\bigg[2\Big(H^L_{-\frac{1}{2},t}H^R_{-\frac{1}{2},t}+H^L_{\frac{1}{2},t}H^R_{\frac{1}{2},t}\Big)
       +2\sin^2{\theta}\Big(H^L_{\frac{1}{2},+}H^R_{\frac{1}{2},+}+H^L_{-\frac{1}{2},-}H^R_{-\frac{1}{2},-}\Big)\nonumber\\
   &  &+2\cos^2{\theta}\Big(H^L_{-\frac{1}{2},0}H^R_{-\frac{1}{2},0}+H^L_{\frac{1}{2},0}H^R_{\frac{1}{2},0}\bigg)
       -2\cos{\theta}\Big(H^L_{-\frac{1}{2},0}H^R_{-\frac{1}{2},t}+H^L_{\frac{1}{2},0}H^R_{\frac{1}{2},t}\nonumber\\
   & & +H^L_{-\frac{1}{2},t}H^R_{-\frac{1}{2},0}+H^L_{\frac{1}{2},t}H^R_{\frac{1}{2},0}\Big)\bigg];\\
 {\cal A}_{VL,SL}^{\rm  int}
   &=&\left(-\frac{2m_\ell}{\sqrt{q^2}}\right)
       \Bigg[H^L_{-\frac{1}{2},t}H^{SPL}_{-\frac{1}{2}}+H^L_{\frac{1}{2},t}H^{SPL}_{\frac{1}{2}}
       -\cos{\theta}\Big(H^L_{-\frac{1}{2},0}H^{SPL}_{-\frac{1}{2}}+H^L_{\frac{1}{2},0}H^{SPL}_{\frac{1}{2}}\Big) \Bigg];\\
 {\cal A}_{VL,SR}^{\rm  int}
   &=&\left(-\frac{2m_\ell}{\sqrt{q^2}}\right)
       \Bigg[H^L_{-\frac{1}{2},t}H^{SPR}_{-\frac{1}{2}}+H^L_{\frac{1}{2},t}H^{SPR}_{\frac{1}{2}}
       -\cos{\theta}\Big(H^L_{-\frac{1}{2},0}H^{SPR}_{-\frac{1}{2}}+H^L_{\frac{1}{2},0}H^{SPR}_{\frac{1}{2}}\Big) \Bigg];\\
  {\cal A}_{VL,T}^{\rm int}
   &=&\left(-\frac{4m_\ell}{\sqrt{q^2}}\right)
     \Bigg\{H^L_{-\frac{1}{2},0}\Big(H^T_{-\frac{1}{2},0,t}-H^T_{-\frac{1}{2},+,-}\Big)
          +H^L_{\frac{1}{2},0}\Big(H^T_{\frac{1}{2},0,t}-H^T_{\frac{1}{2},+,-}\Big)\nonumber\\
   & &+(1+\cos{\theta})\Big(H^L_{\frac{1}{2},+}H^T_{\frac{1}{2},+,t}-H^L_{\frac{1}{2},+}H^T_{\frac{1}{2},+,0}\Big)
      +(1-\cos{\theta})\Big(H^L_{-\frac{1}{2},-}H^T_{-\frac{1}{2},-,t}-H^L_{-\frac{1}{2},-}H^T_{-\frac{1}{2},0,-}\Big) \nonumber\\
   & &+\cos{\theta}\bigg[H^L_{-\frac{1}{2},t}\Big(H^T_{-\frac{1}{2},+,-}-H^T_{-\frac{1}{2},0,t}\Big)
      +H^L_{\frac{1}{2},t}\Big(H^T_{\frac{1}{2},+,-}-H^T_{\frac{1}{2},0,t}\Big)\bigg]\Bigg\};\\
  {\cal A}_{VR,SL}^{\rm int}
   &=&\left(-\frac{2m_\ell}{\sqrt{q^2}}\right)
       \Bigg[H^R_{-\frac{1}{2},t}H^{SPL}_{-\frac{1}{2}}+H^R_{\frac{1}{2},t}H^{SPL}_{\frac{1}{2}}
      -\cos{\theta}\Big(H^R_{-\frac{1}{2},0}H^{SPL}_{-\frac{1}{2}}+H^R_{\frac{1}{2},0}H^{SPL}_{\frac{1}{2}}\Big)\Bigg];\\
  {\cal A}_{VR,SR}^{\rm int}
   &=&\left(-\frac{2m_\ell}{\sqrt{q^2}}\right)
      \Bigg[H^R_{-\frac{1}{2},t}H^{SPR}_{-\frac{1}{2}}+H^R_{\frac{1}{2},t}H^{SPR}_{\frac{1}{2}}
      -\cos{\theta}\Big(H^R_{-\frac{1}{2},0}H^{SPR}_{-\frac{1}{2}}+H^R_{\frac{1}{2},0}H^{SPR}_{\frac{1}{2}}\Big)\Bigg];\\
  {\cal A}^{\rm int}_{VR,T}
   &=&\left(-\frac{4m_\ell}{\sqrt{q^2}}\right)
      \Bigg\{H^R_{\frac{1}{2},0}\Big(H^T_{\frac{1}{2},0,t}-H^T_{\frac{1}{2},+,-}\Big)+ H^R_{-\frac{1}{2},0}\Big(H^T_{-\frac{1}{2},0,t}-H^T_{-\frac{1}{2},+,-}\Big)\nonumber\\
   & &+(1+\cos{\theta})\Big(H^R_{\frac{1}{2},+}H^T_{\frac{1}{2},+,t}-H^R_{\frac{1}{2},+}H^T_{\frac{1}{2},+,0}\Big)+ (1-\cos{\theta})\Big(H^R_{-\frac{1}{2},-}H^T_{-\frac{1}{2},-,t}-H^R_{-\frac{1}{2},-}H^T_{-\frac{1}{2},0,-}\Big)\nonumber\\
   & &+\cos{\theta}\bigg[H^R_{-\frac{1}{2},t}\Big(H^T_{-\frac{1}{2},+,-}-H^T_{-\frac{1}{2},0,t}\Big)
     +H^R_{\frac{1}{2},t}\Big(H^T_{\frac{1}{2},+,-}-H^T_{\frac{1}{2},0,t}\Big)\bigg]\Bigg\};\\
  {\cal A}^{\rm int}_{SL,SR}
  &=&2 \Big(H^{SPL}_{-\frac{1}{2}}H^{SPR}_{-\frac{1}{2}}+H^{SPL}_{\frac{1}{2}}H^{SPR}_{\frac{1}{2}}\Big);\\
   {\cal A}^{\rm int}_{SL,T}
   &=& 4\cos{\theta}\Bigg[H^{SPL}_{-\frac{1}{2}}\Big(H^T_{-\frac{1}{2},+,-}
     -H^T_{-\frac{1}{2},0,t}\Big)+H^{SPL}_{\frac{1}{2}}\Big(H^T_{\frac{1}{2},+,-}-H^T_{\frac{1}{2},0,t}\Big)\Bigg];\\
 {\cal A}^{\rm int}_{SR,T}
   &=& 4\cos{\theta}\Bigg[H^{SPR}_{-\frac{1}{2}}\Big(H^T_{-\frac{1}{2},+,-}
      -H^T_{-\frac{1}{2},0,t}\Big)+H^{SPR}_{\frac{1}{2}}\Big(H^T_{\frac{1}{2},+,-}-H^T_{\frac{1}{2},0,t}\Big)\Bigg].
\end{eqnarray}
In the above equations, we present all possible amplitudes, including interactions between different types of NP operators, even though we assume a single NP operator in current work. In certain specialized NP models in which more than two operators are introduced, the above equations can still be applied directly.

Integrating over $\cos \theta$, one obtains the normalized differential rate,
\begin{eqnarray}
\frac{d\Gamma(\Lambda_c\to \Lambda \ell^{+}\nu_\ell)}{dq^2} =  \frac{G_F^2|V_{cs}|^2q^2\sqrt{Q_+Q_-}}{2^{10}\pi^3M_{\Lambda_c}^3}
\Big(1-\frac{m_\ell^2}{q^2}\Big)^2 \int_{-1}^{1}{\cal A}_{\rm total}d\cos \theta.\label{eq:distr3}
\end{eqnarray}
One then obtains the branching fraction of $\Lambda_c\to \Lambda\ell^{+}\nu_\ell$,
\begin{eqnarray}
\mathcal B(\Lambda_c\to \Lambda\ell^{+}\nu_\ell)=\tau_{\Lambda_c}\int_{m^2}^{M_{-}^2} {\rm d}q^2 \frac{{\rm d}\Gamma(\Lambda_c\to \Lambda\ell^{+}\nu_\ell)}{{\rm d}q^2}\,,\label{brequation}
\end{eqnarray}
where $\tau_{\Lambda_c}$ is the life time of $\Lambda_c$. In order to test the LFU, we also define the differential and integrated ratios as:
\begin{eqnarray}
 &&\mathcal{R}_{\Lambda}(q^2)= \frac{{\rm d}\Gamma(\Lambda_c\to \Lambda\mu^{+}\nu_\mu)/{\rm d}q^2}{{\rm d}\Gamma(\Lambda_c\to \Lambda e^{+}\nu_e)/{\rm d}q^2}\,,\label{eq:rlambda}\\
 &&\mathcal{R}_{\Lambda}= \dfrac{{\displaystyle\int_{m_{\mu}^2}^{M_{-}^2} {\rm d}q^2{\rm d}\Gamma(\Lambda_c\to \Lambda\mu^{+}\nu_\mu)/{\rm d}q^2}}{{\displaystyle\int_{m_{e}^2}^{M_{-}^2} {\rm d}q^2{\rm d}\Gamma(\Lambda_c\to \Lambda e^{+}\nu_e)/{\rm d}q^2}}\,.
\end{eqnarray}

In the experimental side, we often measure the forward-backward asymmetry in the lepton-side, the definition of which is given as
\begin{eqnarray}
  {\cal A}_{\rm FB}(q^2)= \frac{\displaystyle\int_{0}^{1}\frac{d^2\Gamma(\Lambda_c\to \Lambda \ell^{+}\nu_\ell)}{dq^2 d\cos\theta}d \cos\theta-\displaystyle\int_{-1}^{0} \frac{d^2\Gamma(\Lambda_c\to \Lambda \ell^{+}\nu_\ell)}{dq^2 d\cos\theta}d \cos\theta}{\displaystyle\int_{0}^{1}\frac{d^2\Gamma(\Lambda_c\to \Lambda \ell^{+}\nu_\ell)}{dq^2 d\cos\theta}d \cos\theta+\displaystyle\int_{-1}^{0} \frac{d^2\Gamma(\Lambda_c\to \Lambda \ell^{+}\nu_\ell)}{dq^2 d\cos\theta}d \cos\theta}\,.
\end{eqnarray}
In addition, the partial rates $\frac{d\Gamma^{\lambda_2=\frac{1}{2}}}{dq^2}$ and $\frac{d\Gamma^{\lambda_2=-\frac{1}{2}}}{dq^2}$ of $\Lambda_c\to \Lambda \ell^{+}\nu_\ell$ decay for the different helicities ($\lambda_2=\pm \frac{1}{2}$) of the final baryon $\Lambda$ can be calculated, which are shown in Appendix. Based on these,  we could define the $q^2$-dependent longitudinal polarization of  $\Lambda$ as
\begin{eqnarray}
{\cal P}^\Lambda_L(q^2)=\frac{{\rm d}\Gamma^{\lambda_2=\frac{1}{2}}/{\rm d}q^2-
    {\rm d}\Gamma^{\lambda_2=-\frac{1}{2}}/{\rm d}q^2}{{\rm d}\Gamma^{\lambda_2=\frac{1}{2}}/{\rm d}q^2+
    {\rm d}\Gamma^{\lambda_2=-\frac{1}{2}}/{\rm d}q^2}.
\end{eqnarray}
Similarly, we calculate the partial rates for the different helicities ($\lambda_\ell=\pm \frac{1}{2}$) of the final leptons and define the $q^2$-dependent longitudinal polarization of the lepton as
\begin{eqnarray}
{\cal P}^{\ell}_L(q^2)=\frac{{\rm d}\Gamma^{\lambda_\ell=\frac{1}{2}}/{\rm d}q^2-
    {\rm d}\Gamma^{\lambda_\ell=-\frac{1}{2}}/{\rm d}q^2}{{\rm d}\Gamma^{\lambda_\ell=\frac{1}{2}}/{\rm d}q^2+
    {\rm d}\Gamma^{\lambda_\ell=-\frac{1}{2}}/{\rm d}q^2}.
\end{eqnarray}

In order to show the dependence of $\cos^2(\theta)$ in the decay width, a convexity parameter is usually defined as
\begin{eqnarray}\label{CF1}
{\cal C}_F^\ell (q^2) = \frac{1}{ d \Gamma/d q^2} \frac{\,d^2}{d  (\cos\theta )^2 \,}\Bigg(\frac{d^2 \Gamma}{dq^2 d\cos\theta}\Bigg),
\end{eqnarray}
which is measurable in the experiments.

\section{Numerical Results} \label{sec:III}
\subsection{Parameter}
The default values of the input parameters are given as follows:
\begin{gather}\label{para}
M_{\Lambda_c}=2.286\,{\rm GeV}, \,\,
M_{\Lambda}=1.116\, {\rm GeV}, \,\,
M_{n}=0.940\, {\rm GeV}, \,\,
m_{e}=0.511\times 10^{-3}\, {\rm GeV}, \,\,
m_{\mu}=0.106\, {\rm GeV},\nonumber\\
V_{cd}=0.225,\,\,
V_{cs}=0.973,\,\,
G_F=1.166\times 10^{-5}\,{\rm GeV}^{-2}, \,\,
\tau_{\Lambda_c}=0.202\,{\rm ps}.
\end{gather}
Other nonperturbative parameters have been specified in the previous section.

\subsection{Predictions of SM}

With Eq.(\ref{brequation}), we can calculate the branching fractions of $\Lambda_c\to\Lambda\ell^+\nu_{\ell}$ as
\begin{eqnarray}
{\mathcal B}(\Lambda_c\to \Lambda \mu^+\nu_\mu)|_{\rm SM}&=&(3.75\pm0.19)\%;\\
{\mathcal B}(\Lambda_c\to \Lambda e^+\nu_e)|_{\rm SM}&=&(3.88\pm0.19)\%.
\end{eqnarray}
where the uncertainties are all from the form factors. The little differences between our results and those of Ref.~\cite{Meinel:2016dqj} are from the lifetime $\tau_{\Lambda_c}$, and the uncertainties of which are not included in current work. On the experimental side, the branching fractions of $\Lambda_c\to \Lambda \ell^+ \nu_{\ell}$ have been measured by BESIII \cite{BESIII:2022ysa, BESIII:2023jxv}, and the results are
\begin{eqnarray}
    &&\mathcal{B}(\Lambda_c\to\Lambda\mu^+\nu_{\mu})|_{\rm exp}=(3.48\pm0.17\pm0.10)\% ;\\
   & &\mathcal{B}(\Lambda_c\to\Lambda e^+\nu_e)|_{\rm exp}=(3.56\pm0.11\pm0.07)\%.
\end{eqnarray}
A comparison of our theoretical results with experimental data shows that they are generally consistent within the range of both theoretical and experimental uncertainties. However, if the experimental central values are taken seriously, our theoretical predictions tend to be slightly larger than the measured values. This discrepancy underscores the need for further improvements in both experimental measurements and theoretical calculations to achieve greater accuracy and to discern whether potential deviations could signal the presence of NP.

In addition, the ratio $\mathcal{R}_\Lambda^{\mu/e}$ defined as eq.~(\ref{eq:rlambda}) can be calculated in SM as
\begin{eqnarray}
    \mathcal{R}_\Lambda^{\mu/e}|_{\rm SM}=0.968\pm 0.003,
\end{eqnarray}
and it has also been measured as \cite{BESIII:2023jxv}
\begin{align}
   \mathcal{R}_\Lambda^{\mu/e}|_{\rm Exp}=0.98 \pm 0.05_{stat.}\pm 0.03_{syst}.
\end{align}
It can seen that there is little deviation of the SM from the experiment result, though they are consistent within uncertainties.

\begin{table}[!htb]
\begin{center}
\caption{The branching fractions of $\Lambda_c^+ \to \Lambda \ell^+ \nu_\ell $ in units of $\%$.} \label{BRTable}
\begin{tabular}{l|cccc}
\hline
\hline
Model        &\multicolumn{1}{c}{${\cal B}(\Lambda_c^+ \to \Lambda e^+\nu_e) $}  &\multicolumn{1}{c}{${\cal B}(\Lambda_c^+ \to \Lambda {\mu}^+\nu_{\mu}) $ }\\
\hline     LQCD                                  &  $ 3.88 \pm 0.19 $    &     $ 3.75 \pm 0.19$\\
   LQCD~\cite{Meinel:2016dqj} &  $ 3.80 \pm 0.22 $    &     $ 3.69 \pm 0.22 $\\
           HBM\cite{Geng:2022fsr}&  $ 3.78  \pm 0.25 $   &     $ 3.67 \pm 0.23$\\
     CQM~\cite{Gutsche:2015rrt}& $ 2.78 $                  &       $2.69$\\
			RQM~\cite{Faustov:2019ddj}&$ 3.25 $&$3.14 $\\
			QCDSR~\cite{Zhang:2023nxl}&$ 3.49\pm0.65 $&$ 3.37\pm0.54  $\\
			LFQM~\cite{Zhao:2018zcb}&$ 1.63 $&$ - $\\
			LFQM~\cite{Geng:2020gjh} &$ 3.55\pm 0.104 $&$ 3.40\pm 0.102 $\\
			LFQM~\cite{Li:2021qod} &$ 4.04\pm 0.75  $&$ 3.90\pm 0.73 $\\
		   $SU(3)_F$~\cite{Geng:2019bfz,He:2021qnc} & $3.6\pm 0.4 $ & $3.5\pm 0.4$\\
		   Data~\cite{BESIII:2022ysa, BESIII:2023jxv} &$ 3.56\pm 0.13 $&$ 3.48\pm 0.20$\\			\hline
			\hline
					\end{tabular}
\end{center}
	\end{table}
	
	\begin{table}[!htb]
\begin{center}
	\caption{The branching fractions of $\Lambda_c^+ \to n \ell^+\nu_\ell$ in units of $\%$. }
	\label{BRTable2}
	\begin{tabular}{l|cccc}
		\hline
		\hline
		&\multicolumn{1}{c}{${\cal B}(\Lambda_c^+ \to n e^+\nu_e )$}
		&\multicolumn{1}{c}{${\cal B}(\Lambda_c^+ \to n {\mu}^+\nu_{\mu}) $ }\\
		\hline
        LQCD                       &$ 0.418\pm0.029$&$ 0.405 \pm0.029$\\    	
        LQCD~\cite{Meinel:2017ggx} &$ 0.410 \pm0.029$&$ 0.400\pm0.029$\\
        HBM~\cite{Geng:2022fsr}    &$ 0.40 \pm0.04  $&$ 0.40\pm0.04 $\\
		QCDSR~\cite{Zhang:2023nxl} &$ 0.281\pm0.056 $&$ 0.275\pm0.055  $\\
		LFQM~\cite{Zhao:2018zcb}   &$ 0.201 $&$ - $\\
		LFQM~\cite{Geng:2020gjh}   &$ 0.36 \pm0.15 $&$ 0.34 \pm0.15 $\\
        MBM ~\cite{Geng:2020fng}   &$ 0.36 \pm0.15 $&$ 0.34 \pm0.15 $\\
        CCQM~\cite{Gutsche:2014zna}&$ 0.30 $        & \\
        NRQM~\cite{Perez-Marcial:1989sch}&$ 0.28 $        & \\
	   $SU(3)_F$~\cite{He:2021qnc} &$ 0.520 \pm0.046 $&$0.506 \pm0.045 $\\
           $SU(3)_F$~\cite{Lu:2016ogy} &$ 0.293 \pm0.034 $& \\
		\hline
		\hline
		
	\end{tabular}
\end{center}
\end{table}

\begin{figure}[!htb]
\begin{center}
\includegraphics[width=0.45\textwidth]{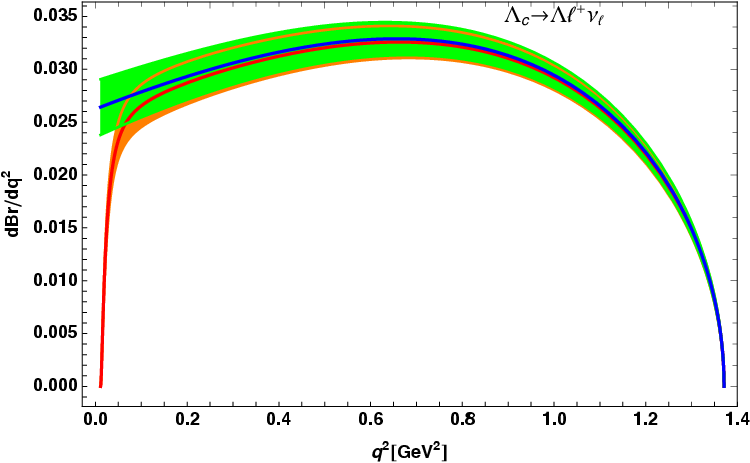}
\includegraphics[width=0.45\textwidth]{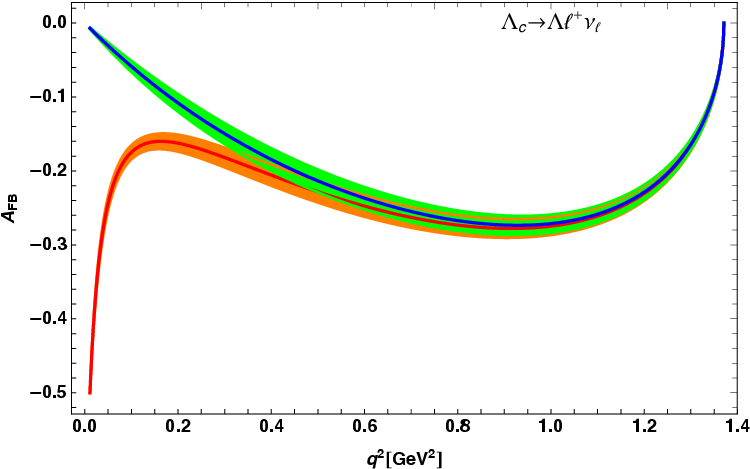}
\includegraphics[width=0.45\textwidth]{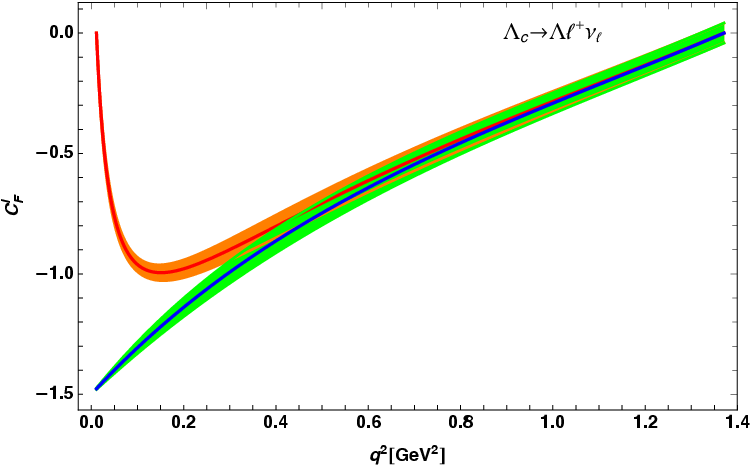}
\includegraphics[width=0.45\textwidth]{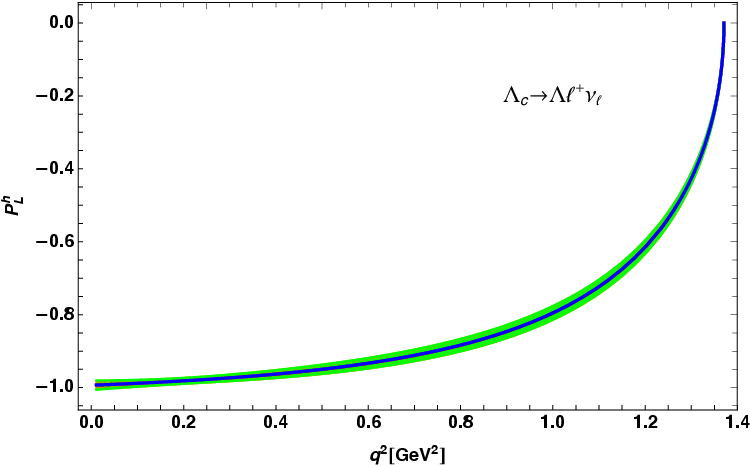}
\includegraphics[width=0.45\textwidth]{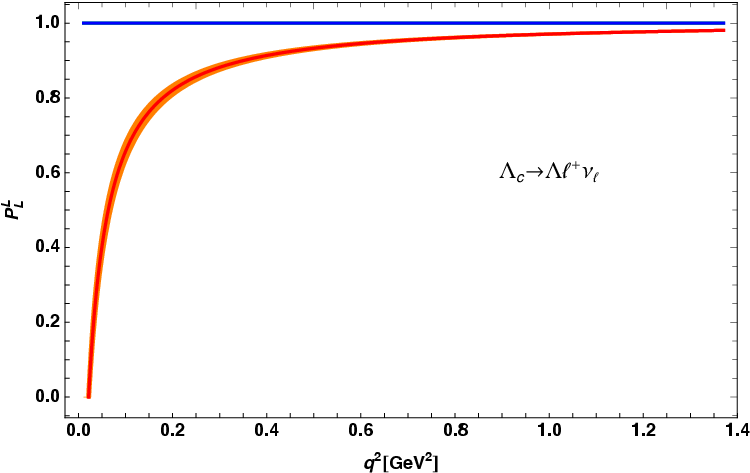}
\caption{The $q^2$-dependence of the differential branching ratios $d{\cal B}/dq^2$, the forward-backward asymmetries on the leptonic side ${\cal A}_{FB}(q^2)$, the convexity parameters ${\cal C}_F^\ell(q^2)$, and the helicity asymmetries of the final baryons and leptons for the decays $\Lambda_c^+ \to \Lambda \ell^+ \nu_\ell$ , respectively.  In all figures, the red lines with orange bands are for muon mode, and the the blue lines with green bands are for electron mode.}
\label{fig:observesmLamf}
\end{center}
\end{figure}

\begin{figure}[!htb]
\begin{center}
\includegraphics[width=0.45\textwidth]{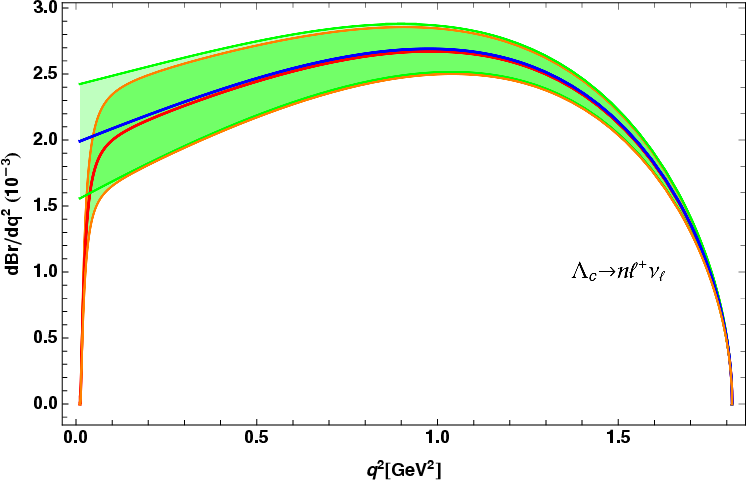}
\includegraphics[width=0.45\textwidth]{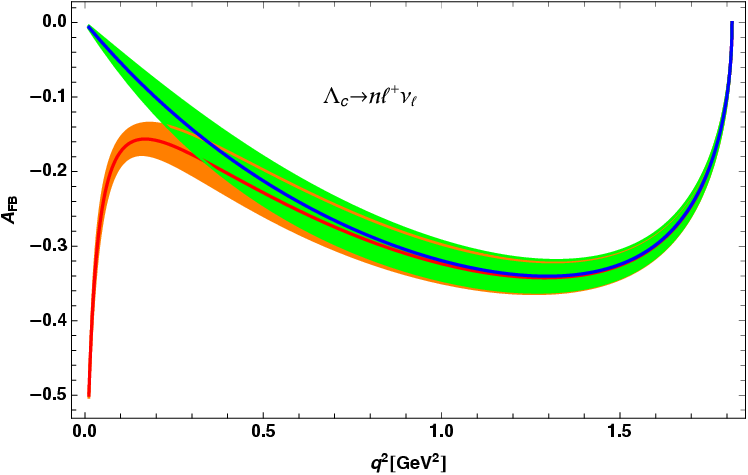}
\includegraphics[width=0.45\textwidth]{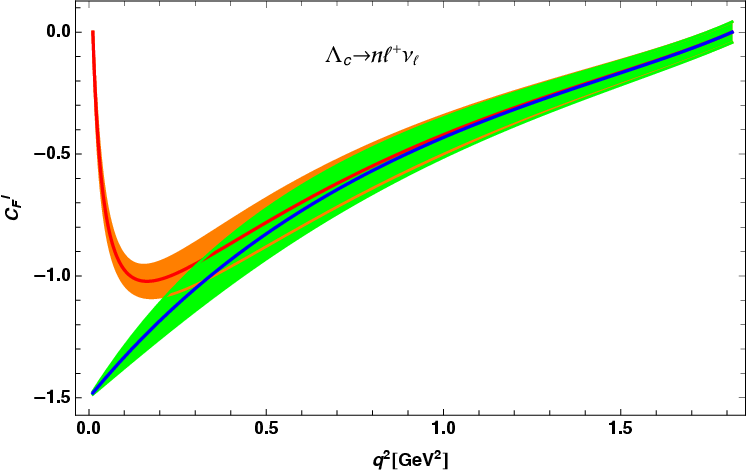}
\includegraphics[width=0.45\textwidth]{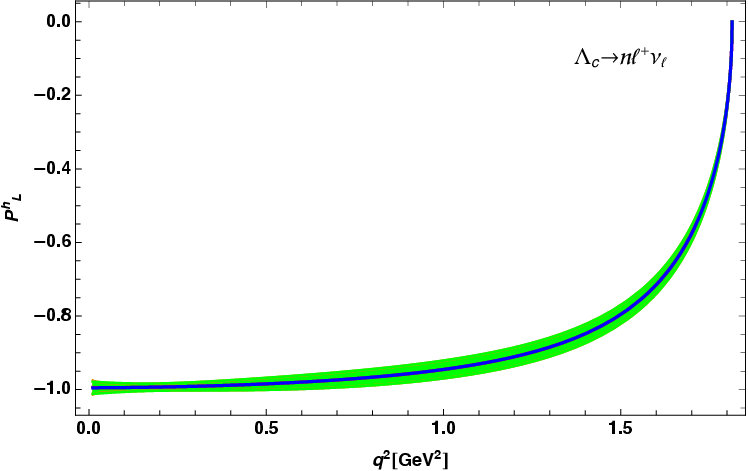}
\includegraphics[width=0.45\textwidth]{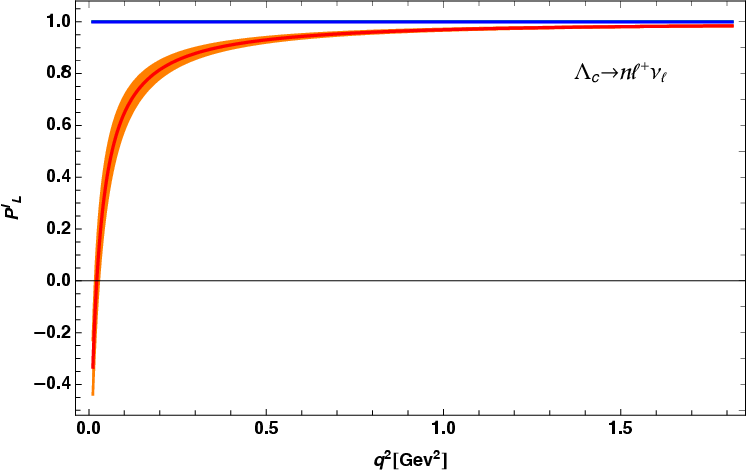}
\caption{The $q^2$-dependence of the differential branching ratios $d{\cal B}/dq^2$, the forward-backward asymmetries on the leptonic side ${\cal A}_{FB}(q^2)$, the convexity parameters ${\cal C}_F^\ell(q^2)$, and the helicity asymmetries of the final baryons and leptons for the decays  $\Lambda_c^+ \to n \ell^+ \nu_\ell$, respectively. In all figures, the red lines with orange bands are for muon modes, and the the blue lines with green bands are for electron modes. }
\label{fig:observesmnf}
\end{center}
\end{figure}

\begin{figure}[!htb]
\begin{center}
\includegraphics[width=0.45\textwidth]{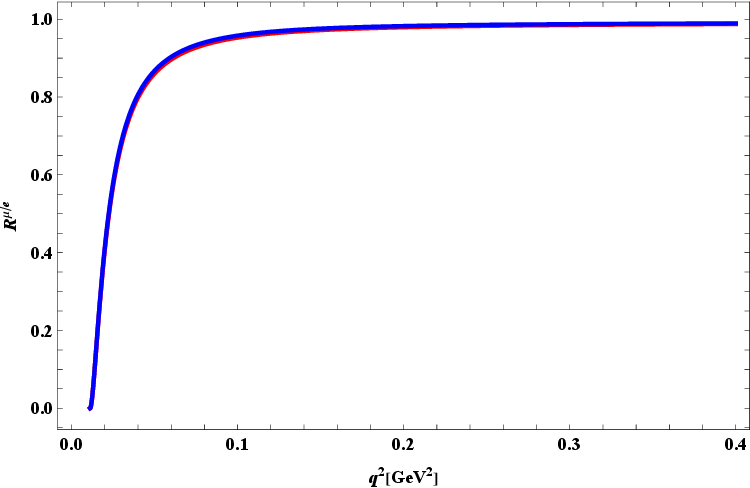}
\includegraphics[width=0.45\textwidth]{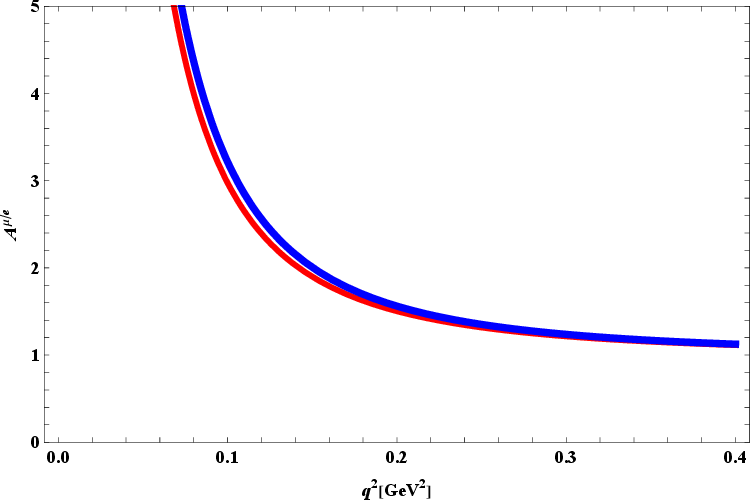}
\caption{The $q^2$-dependence of ratio ${\cal R}_B^{\mu/e}(q^2)$ and  ${\cal A}_B^{\mu/e}(q^2)$, where red line and blue one are for $\Lambda_c^+ \to \Lambda \ell^+ \nu_\ell$ and  $\Lambda_c^+ \to n \ell^+ \nu_\ell$, respectively.}
\label{fig:ReuobserveSM}
\end{center}
\end{figure}

Similarly, the branching fractions of $\Lambda_c\to n \ell^+ \nu_\ell$ in SM are given as
\begin{eqnarray}\label{BFSM}
&&{\mathcal B}(\Lambda_c\to n \mu^+\nu_\mu)|^{\rm SM}=(4.05\pm0.29)\times 10^{-3};\\
&&{\mathcal B}(\Lambda_c\to n e^+\nu_e)|^{\rm SM}=(4.15\pm0.29)\times 10^{-3};
\end{eqnarray}
and the ratio is calculated as
\begin{eqnarray}
    \mathcal{R}_n^{\mu/e}|_{\rm SM}=\frac{\mathcal{B}(\Lambda_c\to n\mu^+\nu_{\mu})}{\mathcal{B}(\Lambda_c\to n e^+\nu_{e})}\Big|_{\rm SM}=0.977\pm0.001.
\end{eqnarray}
However, the experimental results are not available until now.

The theoretical calculations of the branching fractions of $\Lambda_c^+ \to \Lambda \ell^+ \nu_\ell$ based on the homogeneous bag model (HBM)~\cite{Geng:2022fsr}, the covariant quark model (CQM)~\cite{Gutsche:2015rrt}, the relativistic quark model (RQM)~\cite{Faustov:2019ddj}, the lattice QCD \cite{Meinel:2016dqj}, QCD Sum Rules (QCDSR)~\cite{Zhang:2023nxl}, the light-front quark model (LFQM)~\cite{Zhao:2018zcb, Geng:2020gjh, Li:2021qod}  and the $SU(3)_F$ symmetry \cite{He:2021qnc,Geng:2019bfz} are collected in Table.~\ref{BRTable}. Experimental data are also included for comparison. The theoretical branching fractions agree in order of magnitude of experimental measurements. In particular, all theoretical calculations predict ${\cal B}(\Lambda_c^+ \to \Lambda e^+ \nu_e)>{\cal B}(\Lambda_c^+ \to \Lambda \mu^+ \nu_\mu)$; an opposite experimental trend would strongly suggest NP. The results based on $SU(3)_F$ are consistent with the experimental data, as they use experimental values as input. The LFQM branching fraction for $\Lambda_c^+ \to \Lambda e^+\nu_e$ from Ref.~\cite{Zhao:2018zcb} is significantly smaller than the experimental value. Although Refs.~\cite{Geng:2020gjh, Li:2021qod} demonstrate that the LFQM can reproduce experimental data with different parameter sets, this raises concerns about the model's predictive power. In Ref.~\cite{Geng:2022fsr}, using the HBM with parameters fitted to mass spectra, the calculated branching fractions have central values slightly larger than the experimental data but smaller than our predictions.

We also present the theoretical predictions for the branching fractions of $\Lambda_c^+ \to n \ell^+ \nu_\ell$ based on lattice QCD (LQCD)\cite{Meinel:2017ggx}, the heavy baryon model (HBM)\cite{Geng:2022fsr}, QCD sum rules (QCDSR)\cite{Zhang:2023nxl}, the light-front quark model (LFQM)\cite{Zhao:2018zcb, Geng:2020gjh}, the MIT bag model (MBM)\cite{Geng:2020fng}, the covariant confined quark model (CCQM)\cite{Gutsche:2014zna}, the non-relativistic quark model (NRQM)\cite{Perez-Marcial:1989sch}, and $SU(3)_F$ symmetry\cite{He:2021qnc, Lu:2016ogy}, as summarized in Table~\ref{BRTable2}. By comparison, we find that our results are consistent with those from HBM~\cite{Geng:2022fsr}, LFQM~\cite{Geng:2020gjh}, and MBM~\cite{Geng:2020fng}. However, our results are larger than those from QCDSR~\cite{Zhang:2023nxl}, CCQM~\cite{Gutsche:2014zna}, and NRQM~\cite{Perez-Marcial:1989sch}, while being smaller than the predictions in \cite{He:2021qnc}. It should be noted that, although LFQM was used in both studies, the results in \cite{Zhao:2018zcb} are approximately half of those in \cite{Geng:2020gjh} due to differences in the wave functions of $\Lambda_c$ adopted in the respective approaches. We also observe that the branching fractions in \cite{He:2021qnc} are about twice those in \cite{Lu:2016ogy}, as $SU(3)_F$ symmetry-breaking effects were included in \cite{He:2021qnc}. On the experimental side, neither $\Lambda_c^+ \to n e^+ \nu_e$ nor $\Lambda_c^+ \to n \mu^+ \nu_\mu$ has been measured to date. Future high-precision measurements will provide valuable insights to test and differentiate between these theoretical approaches.

In Figures.~\ref{fig:observesmLamf} and \ref{fig:observesmnf}, we show the $q^2$ dependence of the differential branching ratios $d{\cal B}/dq^2$, the forward-backward asymmetries on the leptonic side ${\cal A}_{FB}(q^2)$, the convexity parameters ${\cal C}_F^\ell(q^2)$, and the helicity asymmetries of the final baryons and leptons for the decays $\Lambda_c^+ \to \Lambda \ell^+ \nu_\ell$ and $\Lambda_c^+ \to n \ell^+ \nu_\ell$, respectively. From these figures, it is clear that the plots for decays involving electrons and muons nearly coincide near the zero-recoil point, $q^2 = q^2_{\rm max} = (M_{\Lambda_c} - M_{\Lambda(n)})^2$. However, the differential decay rates, the forward-backward asymmetry ${\cal A}_{FB}$, and the convexity parameter ${\cal C}_F^\ell$ exhibit markedly different behaviors near the maximum recoil point, $q^2 = q_{\rm min}^2 = m_\ell^2$. Specifically, the forward-backward asymmetry ${\cal A}_{FB}$ approaches zero for the decay $\Lambda_c^+ \to \Lambda(n) e^+ \nu_e$ as $q^2 \to 0$, while it tends to $-0.5$ for the decay $\Lambda_c^+ \to \Lambda(n) \mu^+ \nu_\mu$. Similarly, the convexity parameter ${\cal C}_F^\ell$ approaches $-1.5$ for the electron mode and $0$ for the muon mode at $q^2 = q_{\rm min}^2$.  Additionally, we observe that the $q^2$-dependent longitudinal polarization of the final baryons is nearly identical across the entire kinematic range for both decay modes. For the longitudinal polarization of the charged leptons, the electron case reflects the chiral limit of a massless lepton, where the lepton is purely left-handed. In contrast, for the muon mode, the helicity asymmetry decreases from 1 to a negative value of $-0.32$ at zero recoil. Furthermore, it is evident that the longitudinal polarizations of the final baryons and charged leptons are less sensitive to uncertainties in the form factors, as these uncertainties cancel between the numerator and denominator. Therefore, the observables ${\cal P}_L^h$ and ${\cal P}_L^\ell$ are good probes for searching for the effects of new physics beyond SM.

To investigate LFV, in addition to the observables, we introduce another physical parameter defined as:
\begin{eqnarray}
\mathcal{A}_{B}^{\mu/e}(q^2) = \frac{\mathcal{A}_{\rm FB}(q^2)(\Lambda_c^+ \to B \mu^+ \nu_\mu)}{\mathcal{A}_{\rm FB}(q^2)(\Lambda_c^+ \to B e^+ \nu_e)},
\label{eq:Alambda}
\end{eqnarray}
where $B$ denotes either $\Lambda$ or $n$. In SM, we present the $q^2$-dependence of $\mathcal{R}_B^{\mu/e}(q^2)$ and $\mathcal{A}_{B}^{\mu/e}(q^2)$ for $q^2 \in [0,0.4] {\rm GeV}^2$  in Figure.~\ref{fig:ReuobserveSM}. In this figure, the blue lines correspond to the decay $\Lambda_c^+ \to \mu \ell^+ \nu_\ell$, while the red lines represent $\Lambda_c^+ \to  n \ell^+ \nu_\ell$. Both observables exhibit minimal theoretical uncertainties due to the cancellation of many uncertainties in the numerator and denominator. Specifically, when $q^2 \in [0,0.2] {\rm GeV}^2$, $\mathcal{R}_B^{\mu/e}(q^2)$ and $\mathcal{A}_{B}^{\mu/e}(q^2)$ show distinctive trends. As $q^2$ increases beyond $0.2~{\rm GeV}^2$, $\mathcal{R}_B^{\mu/e}(q^2)$ converges to approximately 0.986, while $\mathcal{A}_{B}^{\mu/e}(q^2)$ approaches 1. We also emphasize that the theoretical uncertainties are significantly smaller for $\mathcal{A}_{B}^{\mu/e}(q^2)$, as the input parameter uncertainties cancel out twice in its calculation, which makes $\mathcal{A}_{B}^{\mu/e}(q^2)$ an especially robust observable. Therefore, precise future measurements of $\mathcal{A}_{B}^{\mu/e}(q^2)$ have the potential to provide crucial insights and further enhance the search for LFV.

\subsection{The Model Independent Analysis} \label{subsect:4}
\begin{figure}[!htb]
\begin{center}
\includegraphics[width=0.45\textwidth]{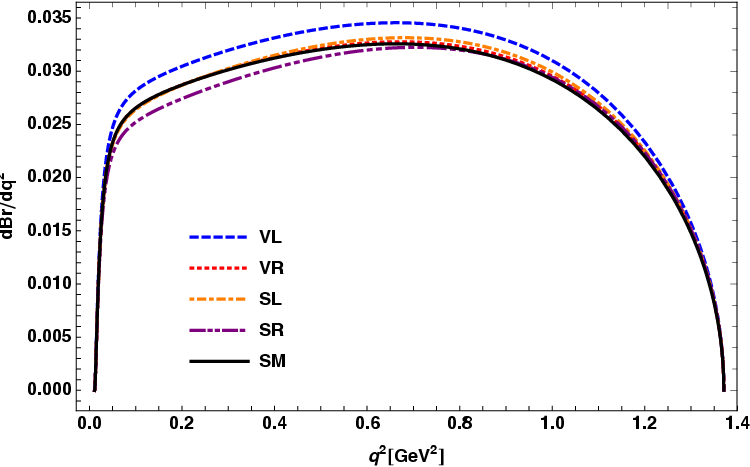}
\includegraphics[width=0.45\textwidth]{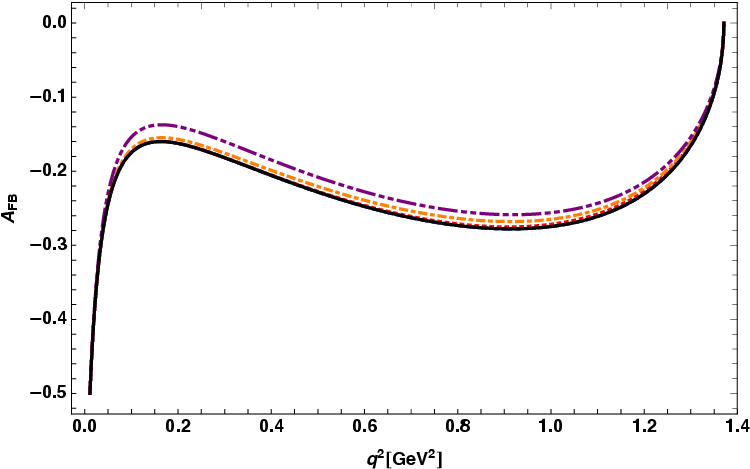}
\includegraphics[width=0.45\textwidth]{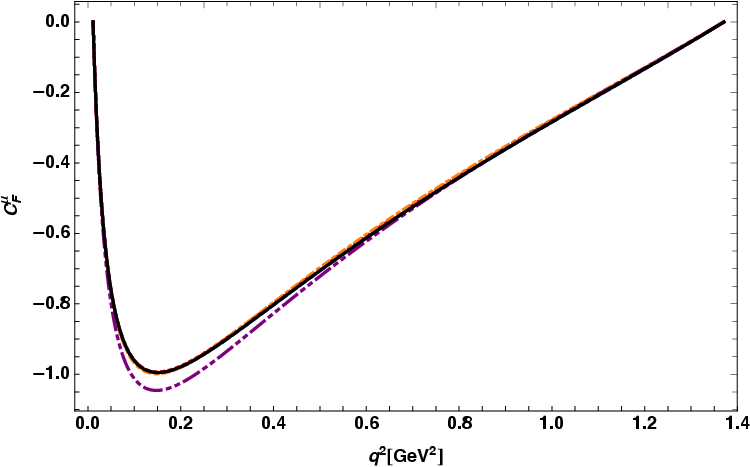}
\includegraphics[width=0.45\textwidth]{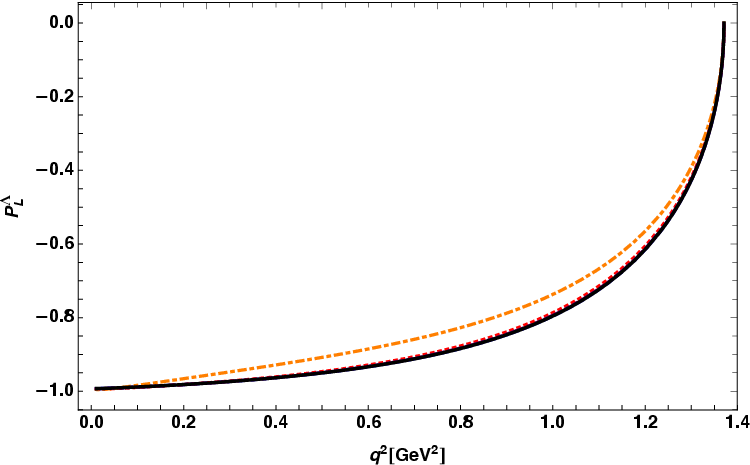}
\includegraphics[width=0.45\textwidth]{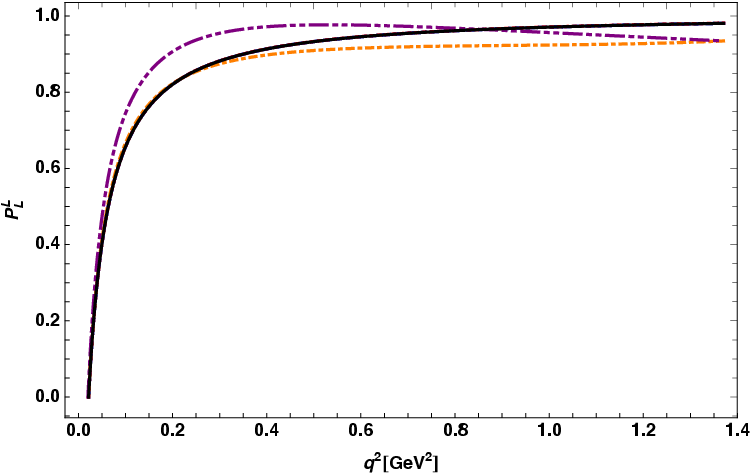}
\includegraphics[width=0.45\textwidth]{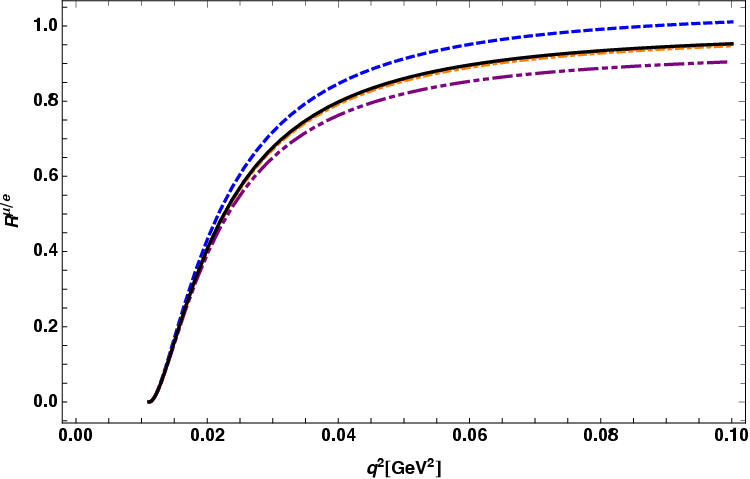}
\includegraphics[width=0.45\textwidth]{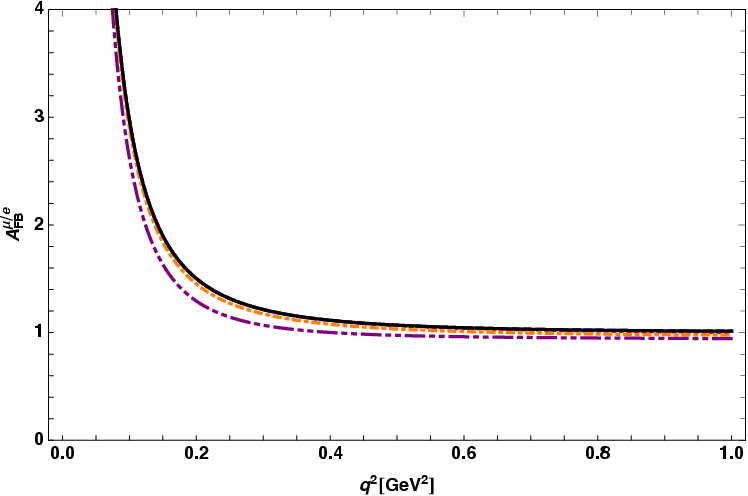}
\caption{The $q^2$-dependence of differential ratios $dBr/dq^2$, the R value, the forward-backward asymmetries of the leptonic side ${\cal A}_{FB}(q^2)$, the convexity parameters ${\cal C}_F^\ell(q^2)$, and the transverse polarization components of the $\Lambda$ and leptons of $\Lambda_c^+ \to \Lambda \mu^+ \nu_\mu$ with NP operators. }
\label{fig:Lambdac_lambda_NP}
\end{center}
\end{figure}

\begin{figure}[!htb]
\begin{center}
\includegraphics[width=0.4\textwidth]{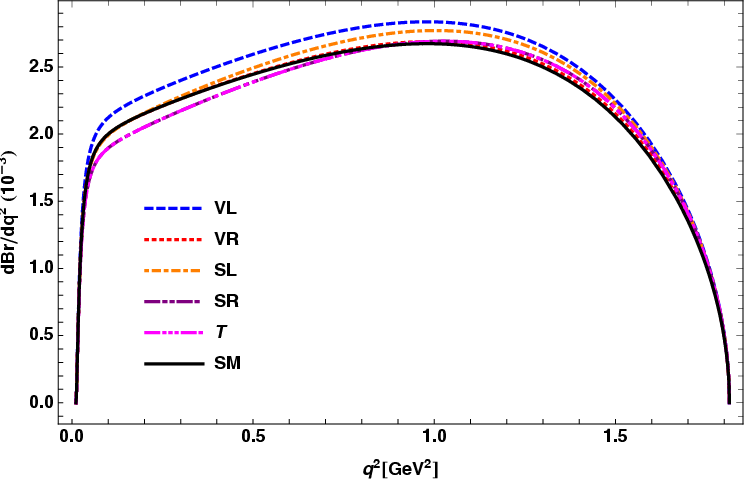}
\includegraphics[width=0.4\textwidth]{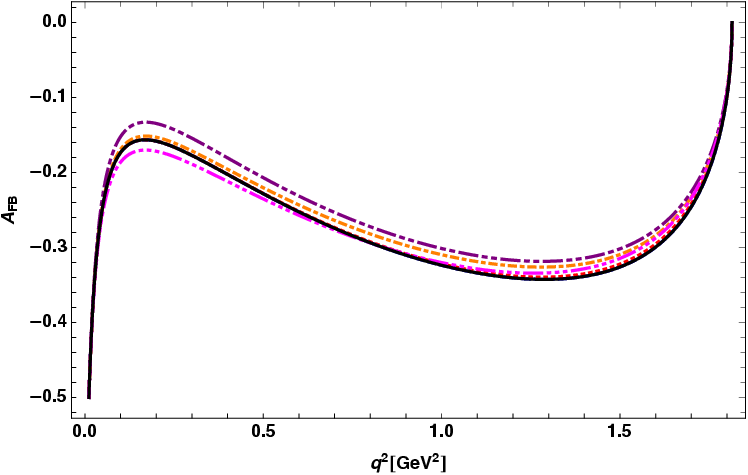}
\includegraphics[width=0.4\textwidth]{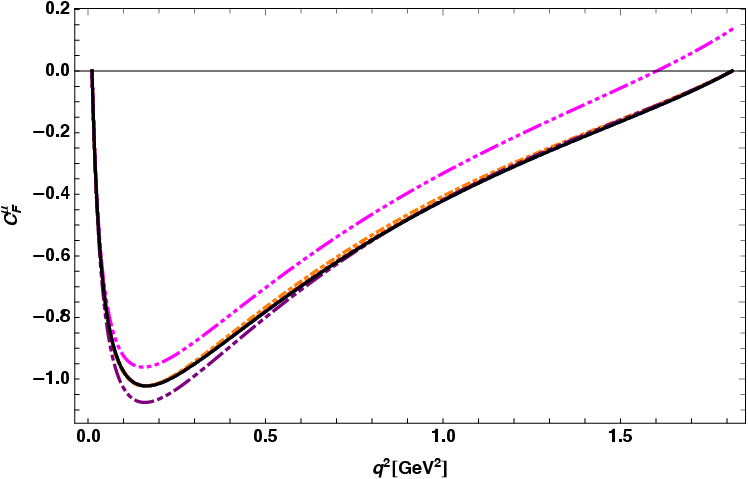}
\includegraphics[width=0.4\textwidth]{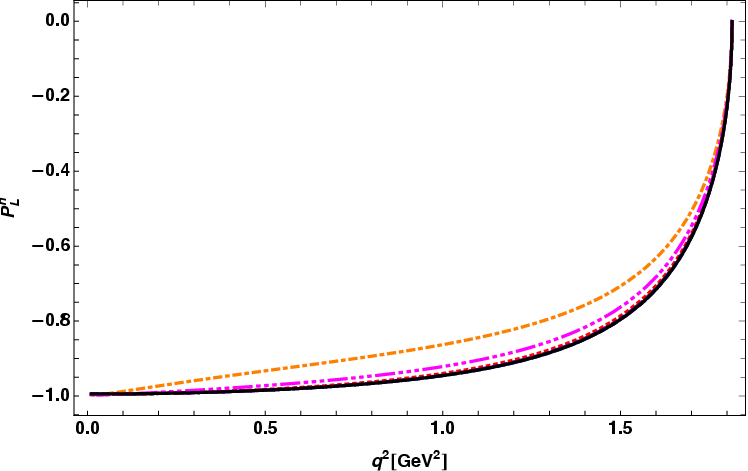}
\includegraphics[width=0.4\textwidth]{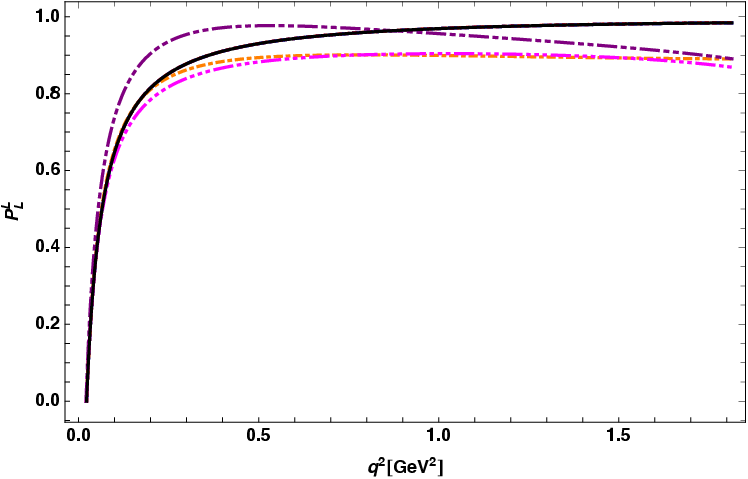}
\includegraphics[width=0.4\textwidth]{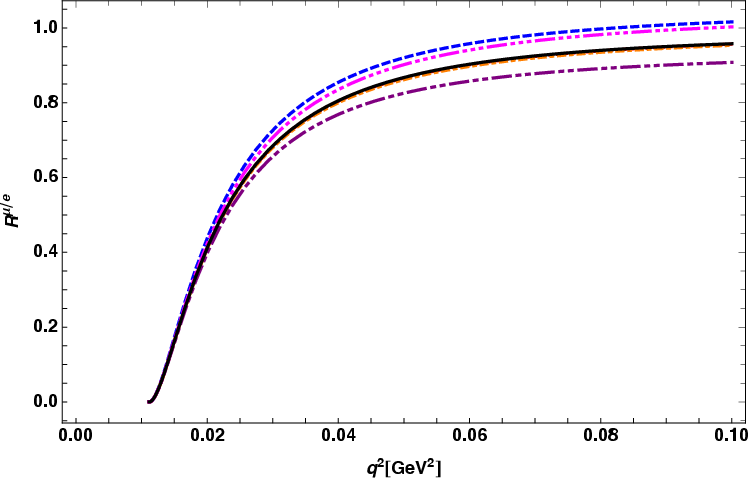}
\includegraphics[width=0.4\textwidth]{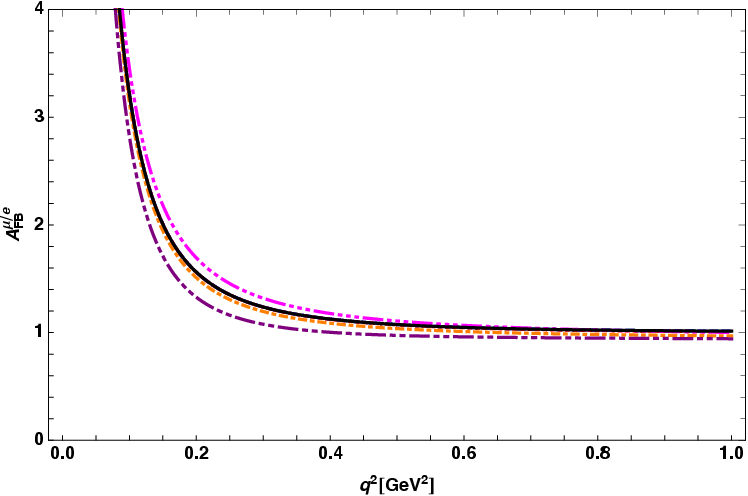}
\caption{The $q^2$-dependence of differential ratios $dBr/dq^2$, the $\cal R$ value, the forward-backward asymmetries of the leptonic side ${\cal A}_{FB}(q^2)$, the convexity parameters ${\cal C}_F^\ell(q^2)$, and the transverse polarization components of the $n$ and leptons of $\Lambda_c^+ \to n \mu^+ \nu_\mu$ with NP operators in NP. }
\label{fig:Lambdac_n_NP}
\end{center}
\end{figure}

The decays $\Lambda_c^+ \to \Lambda \ell^+ \nu_\ell$ arise from the $c \to s \ell^+ \nu_\ell$ transitions, which similarly drive the decays of $D$ mesons. These include pure-leptonic decays such as $D_s^+ \to \ell \nu_\ell$ and semi-leptonic decays like $D \to K^{(*)} \ell^+ \nu_\ell$ and $D_s^+ \to \phi \ell^+ \nu_\ell$. Likewise, the $c \to d \ell^+ \nu_\ell$ transitions govern the decays $\Lambda_c^+ \to n \ell^+ \nu_\ell$, $D^+ \to \ell \nu_\ell$, $D \to \rho (\pi) \ell^+ \nu_\ell$, and $D_s \to K^{(*)} \ell^+ \nu_\ell$. In contrast to the decays of $\Lambda_c$, the decays of $D$ mesons have been measured with high precision in the BESIII and Belle experiments \cite{ParticleDataGroup:2024cfk}. While there are significant uncertainties in both theoretical calculations and experimental measurements, the current experimental results are broadly consistent with the Standard Model predictions. However, the possibility of new physics cannot be definitively ruled out. By utilizing the available experimental data, constraints can be placed on the Wilson coefficients associated with operators of NP, especially within the framework of a single-operator analysis \cite{Wang:2014uiz, Fajfer:2015ixa, Soni:2018adu, Fleischer:2019wlx, Leng:2020fei, Faustov:2019mqr}. For the purposes of this discussion, we restrict our attention to scenarios in which new physics affects only the muon sector. In Ref.\cite{Leng:2020fei}, the authors performed a minimum $\chi^2$ fit to the Wilson coefficients of each operator, based on the latest experimental data. They found that the coefficients $C_{VL}$ and $C_{VR}$ are of order ${\cal O}(10^{-3})$, while $C_{SL}$ and $C_{SR}$ could be of order ${\cal O}(10^{-2})$. However, Ref.\cite{Becirevic:2020rzi} suggests that $C_{VL(R)}$ and $C_{SL(R)}$ could be as large as ${\cal O}(10^{-2})$ and ${\cal O}(10^{-1})$, respectively. More recently, the authors of Ref.\cite{Bolognani:2024cmr} considered complex Wilson coefficients and obtained similar results.  Based on above results, in order to maximize the manifestation of new physics effects while maintaining generality, we adopt
\begin{align}
   C_{VL}=0.03, C_{VR}=-0.01,  C_{SL}=0.3, C_{SR}=0.3, C_{T}=0.15.
\end{align}
Using above Wilson coefficients, we plot the $q^2$ dependence of the differential branching ratios $d{\cal B}/dq^2$, the forward-backward asymmetries on the leptonic side ${\cal A}_{FB}(q^2)$, the convexity parameters ${\cal C}_F^\ell(q^2)$, and the helicity asymmetries of the final baryons and leptons, and lepton flavor violation parameters ${\cal R}^{\mu/e}(q^2)$ and ${\cal A}_{FB}^{\mu/e}(q^2)$  for the decays $\Lambda_c^+ \to \Lambda \mu^+ \nu_\mu$ and $\Lambda_c^+ \to n \mu^+ \nu_\mu$ in Figure.~\ref{fig:Lambdac_lambda_NP} and Figure.~\ref{fig:Lambdac_n_NP}, respectively.  For the decay process $\Lambda_c^+ \to \Lambda \mu^+ \nu_\mu$, the contribution of the tensor operator ($O_T$) is not considered, as the form factors for the matrix element $\langle \Lambda | \bar{s} i \sigma^{\mu\nu} c | \Lambda_c \rangle$ are not yet available.

Figure~\ref{fig:Lambdac_lambda_NP} illustrates that for the decay process $\Lambda_c^+ \to \Lambda \mu^+ \nu_\mu$, the differential decay branching fraction can be significantly enhanced by the left-handed vector operator $O_{VL}$ or suppressed by the right-handed scalar operator $O_{SR}$. The total branching fraction may vary by approximately $10\% $ as a result. Notably, the effects of $O_{VL}$ and $O_{SR}$ on the observable ${\cal R}_{\Lambda_c}^{\mu/e}(q^2)$ are pronounced, particularly in the large $q^2$ region. In contrast, the forward-backward asymmetry of the leptonic side, ${\cal A}_{FB}(q^2)$, as well as the convexity parameters ${\cal C}_F^\ell(q^2)$ and the ratio ${\cal A}_{FB}^{\mu/e}(q^2)$, remain almost unaffected, though the left-handed scalar operator $O_{SR}$ introduces slight shape modifications. The polarization parameters ${\cal P}_L^{\Lambda_c}$ and ${\cal P}_L^{\mu}$, on the other hand, are influenced by scalar operators due to the potential for helicity flipping. Since helicity flipping is proportional to the fermion mass and neutrinos are inherently left-handed, the longitudinal polarization of the final-state baryon, ${\cal P}_L^{\Lambda_c}$, is significantly impacted by the operator $O_{SR}$. Meanwhile, ${\cal P}_L^{\mu}$ is affected by both $O_{SL}$ and $O_{SR}$. These polarization parameters are particularly useful for probing the presence of scalar operators, as they exhibit minimal sensitivity to other operators. Interestingly, when comparing theoretical predictions with experimental data, the prediction of SM appears to slightly overestimate the experimental result, implying that the right-handed scalar operator $O_{SR}$ with a negative Wilson coefficient is much favored.

In Figure~\ref{fig:Lambdac_n_NP}, it is evident that the decay $\Lambda_c^+ \to n \mu^+ \nu_\mu$ exhibits a behavior similar to that of $\Lambda_c^+ \to \Lambda \mu^+ \nu_\mu$. The tensor form factors for $\Lambda_c^+ \to n$ have been calculated in the lattice QCD \cite{Meinel:2017ggx}, and the contributions of the tensor operator are incorporated in Figure~\ref{fig:Lambdac_n_NP}. Although the inclusion of the tensor operator can enhance the total branching fraction by about $10\%$, the differential decay branching fraction with the tensor operator is smaller than that predicted by SM in the low $q^2$ region and larger in the high $q^2$ region. Furthermore, while the tensor operator can suppress ${\cal P}_L^{\mu}$, its effect cannot be distinguished from that of $O_{SL}$. However, the convexity parameter ${\cal C}_F^\ell(q^2)$ is highly sensitive to the tensor operator, suggesting that it could serve as a useful probe to test the impact of the tensor operator.

%

\section{Summary}\label{sec:4}
The study of $\Lambda_c$ decays offers a valuable opportunity to probe new physics (NP) while simultaneously testing the parameters of the Standard Model (SM). In this work, we analyze the decays $\Lambda_c \to (\Lambda, n) \ell^+ \nu_\ell$ ($\ell = \mu, e$) within a model-independent framework. In the absence of right-handed neutrinos, we calculate the helicity amplitudes in detail, considering all possible four-fermion operators, including those involving interactions between different NP operators. For the form factors of $\Lambda_c \to (\Lambda, n)$, we adopt results from lattice QCD calculations. For the branching fractions within the SM, our results for $\Lambda_c \to \Lambda \ell^+ \nu_\ell$ are consistent with previous studies and in agreement with current experimental data, though the central values of our results are approximately $10\%$ larger than the experimental measurements. The decays $\Lambda_c \to n \ell^+ \nu_\ell$ are currently being measured by BESIII. Furthermore, we investigate the $q^2$-dependence of several physical observables, including differential branching fractions, lepton flavor universality ratios, lepton polarization asymmetries, longitudinal helicity fractions of the final baryons, forward-backward asymmetries ${\cal A}_{FB}$, and the convexity parameters ${\cal C}_F^\ell(q^2)$ associated with $\Lambda_c \to (\Lambda, n) \ell^+ \nu\ell$ decays.

Assuming that NP particles couple exclusively to the muon, the Wilson coefficients of each NP operator had been fitted using available data on $D$ and $D_s$ meson decays in the literature. With these fitted Wilson coefficients, we explore the NP effects in $\Lambda_c \to (\Lambda, n) \mu^+ \nu_\mu$ decays to the largest extend. Our results indicate that most physical observables are not sensitive to NP effects, and most NP contributions would likely be obscured by theoretical and experimental uncertainties. Specifically, the differential branching fraction of $\Lambda_c^+ \to \Lambda \mu^+ \nu_\mu$ can be enhanced by the left-handed vector operator $O_{VL}$ or suppressed by the right-handed scalar operator $O_{SR}$ by up to $10\%$. We also note that the ratio of the forward-backward asymmetry in the $\Lambda_c \to (\Lambda, n) \mu^+ \nu_\mu$ decay to that in the $\Lambda_c \to (\Lambda, n) e^+ \nu_e$ decay is less sensitive to hadronic uncertainties and is largely unaffected by the current NP operators. A significant deviation between the experimental data and theoretical predictions could provide a signature for the presence of NP. All of our theoretical results are testable in current experiments, including BESIII, Belle-II, and LHCb, as well as future high-energy experiments such as the Super Tau-Charm Factory and the Circular Electron Positron Collider (CEPC).

\section*{Note Added}
After we completed the manuscript, the BESIII Collaboration reported the first observation of the Cabibbo-suppressed decay $\Lambda_c \to n e^+ \nu_e$ in Ref.\cite{BESIII:2024mgg}, based on $4.5~{\rm fb}^{-1}$ of electron-positron annihilation data. The absolute branching fraction was measured to be $(3.57 \pm 0.34_{\rm stat.} \pm 0.14_{\rm syst.})\times 10^{-3}$, which is consistent with our Standard Model (SM) prediction within uncertainties. Interestingly, the central value of the measurement is slightly lower than our prediction, a trend also observed in the case of $\Lambda_c \to \Lambda e^+ \nu_e$. Since our analysis assumes that new physics (NP) contributions affect only the muon sector, this result does not alter our conclusions.

\section*{Acknowledgments}
This work is supported in part by the National Science Foundation of China under the Grants No. 11925506, 12375089, 12435004, and the Natural Science Foundation of Shandong province under the Grant No. ZR2022ZD26.

\begin{appendix}

\section{The Helicity-dependent Differential Decay Rates}
We can write the total amplitude as
\begin{align}
    \frac{d\Gamma}{dq^2}= \frac{d\Gamma^{\lambda_2=\frac{1}{2}}}{dq^2}+ \frac{d\Gamma^{\lambda_2=-\frac{1}{2}}}{dq^2}= \frac{d\Gamma^{\lambda_\ell=\frac{1}{2}}}{dq^2}+ \frac{d\Gamma^{\lambda_\ell=-\frac{1}{2}}}{dq^2},
\end{align}
and
\begin{align}
\frac{d\Gamma^{\lambda_2}}{dq^2}=&\frac{8}{3}N \bigg[|1+C_{VL}|^2B_{VL}^{\lambda_2}+|C_{VR}|^2B_{VR}^{\lambda_2}+|C_{SL}|^2B_{SL}^{\lambda_2}+|C_{SR}|^2B_{SR}^{\lambda_2}+|C_T|^2B_T^{\lambda_2}\nonumber\\
  &+2Re[(1+C_{VL})^*C_{VR}]B^{\lambda_2}_{VL,VR}+2Re[(1+C_{VL})^*C_{SL}]B^{\lambda_2}_{VL,SL}\nonumber\\
    &+2Re[(1+C_{VL})^*C_{SR}]B^{\lambda_2}_{VL,SR}+2Re[(1+C_{VL})^*C_T]B^{\lambda_2}_{VL,T}\nonumber\\
    &+2Re[C_{VR}^*C_{SL}]B^{\lambda_2}_{VR,SL}+2Re[C_{VR}^*C_{SR}]B^{\lambda_2}_{VR,SR}+2Re[C_{VR}^*C_T]B^{\lambda_2}_{VR,T}\nonumber\\
    &+2Re[C_{SL}^*C_{SR}]B^{\lambda_2}_{SL,SR}+2Re[C_{SL}^*C_T]B^{\lambda_2}_{SL,T}+2Re[C_{SR}^*C_T]B^{\lambda_2}_{SR,T}\bigg],
\end{align}
with
\begin{align}
    N=\frac{G_F^2|V_{cs(d)}|^2q^2\sqrt{Q_+Q_-}}{1024\pi^3M_{\Lambda_c}^3}\Big(1-\frac{m_\ell^2}{q^2}\Big)^2.
\end{align}
The inner functions are given as
\begin{align}
&B^{\lambda_2=\frac{1}{2}}_{VL}=|H^L_{\frac{1}{2},+}|^2+|H^L_{\frac{1}{2},0}|^2+\frac{m_\ell^2}{2q^2}\Big(3|H^L_{\frac{1}{2},t}|^2+|H^L_{\frac{1}{2},+}|^2+|H^L_{\frac{1}{2},0}|^2\Big);\\
    &B^{\lambda_2=\frac{1}{2}}_{VR}=|H^R_{\frac{1}{2},+}|^2+|H^R_{\frac{1}{2},0}|^2+\frac{m_\ell^2}{2q^2}\Big(3|H^R_{\frac{1}{2},t}|^2+|H^R_{\frac{1}{2},+}|^2+|H^R_{\frac{1}{2},0}|^2\Big);\\
    &B^{\lambda_2=\frac{1}{2}}_{SL}=\frac{3}{2}|H^{SPL}_{\frac{1}{2}}|^2;\\
    &B^{\lambda_2=\frac{1}{2}}_{SR}=\frac{3}{2}|H^{SPR}_{\frac{1}{2}}|^2;\\
    &B^{\lambda_2=\frac{1}{2}}_{T}=2\bigg[\Big(H^T_{\frac{1}{2},+,t}-H^T_{\frac{1}{2},+,0}\Big)^2+(H^T_{\frac{1}{2},+,-}-H^T_{\frac{1}{2},0,t}\Big)^2\bigg]\nonumber\\
    &\,\,\,\,\,\,\,\,\,\,\,\,\,\,\,\,\,\,\,\,\,\,\,\,\,\,\,\,\,\,
    +\frac{4m^2}{q^2}\bigg[\Big(H^T_{\frac{1}{2},+,t}-H^T_{\frac{1}{2},+,0}\Big)^2+\Big(H^T_{\frac{1}{2},+,-}-H^T_{\frac{1}{2},0,t}\Big)^2\bigg];\\
    &B^{\lambda_2=\frac{1}{2}}_{VL,VR}=H^L_{\frac{1}{2},+}H^R_{\frac{1}{2},+}+H^L_{\frac{1}{2},0}H^R_{\frac{1}{2},0}+\frac{m_\ell^2}{2q^2}\Big(3H^L_{\frac{1}{2},t}H^R_{\frac{1}{2},t}+H^L_{\frac{1}{2},+}H^R_{\frac{1}{2},+}+H^L_{\frac{1}{2},0}H^R_{\frac{1}{2},0}\Big);\\
    &B^{\lambda_2=\frac{1}{2}}_{VL,SL}=-\frac{3m_\ell}{2\sqrt{q^2}}H^L_{\frac{1}{2},t}H^{SPL}_{\frac{1}{2}};\\
    &B^{\lambda_2=\frac{1}{2}}_{VL,SR}=-\frac{3m_\ell}{2\sqrt{q^2}}H^L_{\frac{1}{2},t}H^{SPR}_{\frac{1}{2}};\\
    &B^{\lambda_2=\frac{1}{2}}_{VL,T}=-\frac{3m_\ell}{\sqrt{q^2}}\Big[H^L_{\frac{1}{2},+}\Big(H^T_{\frac{1}{2},+,t}-H^T_{\frac{1}{2},+,0}\Big)+H^L_{\frac{1}{2},0}\Big(H^T_{\frac{1}{2},0,t}-H^T_{\frac{1}{2},+,-}\Big)\Big];\\
     &B^{\lambda_2=\frac{1}{2}}_{VR,SL}=-\frac{3m_\ell}{2\sqrt{q^2}}H^R_{\frac{1}{2},t}H^{SPL}_{\frac{1}{2}};\\
    &B^{\lambda_2=\frac{1}{2}}_{VR,SR}=-\frac{3m_\ell}{2\sqrt{q^2}}H^R_{\frac{1}{2},t}H^{SPR}_{\frac{1}{2}};\\
    &B^{\lambda_2=\frac{1}{2}}_{VR,T}=-\frac{3m_\ell}{\sqrt{q^2}}\Big[H^R_{\frac{1}{2},+}\Big(H^T_{\frac{1}{2},+,t}-H^T_{\frac{1}{2},+,0}\Big)+H^R_{\frac{1}{2},0}\Big(H^T_{\frac{1}{2},0,t}-H^T_{\frac{1}{2},+,-}\Big)\Big];\\
    &B^{\lambda_2=\frac{1}{2}}_{SL,SR}=\frac{3}{2}H^{SPL}_{\frac{1}{2}}H^{SPR}_{\frac{1}{2}};\\
    &B^{\lambda_2=\frac{1}{2}}_{SL,T}=0;\\
    &B^{\lambda_2=\frac{1}{2}}_{SR,T}=0;\\
     &B^{\lambda_2=-\frac{1}{2}}_{VL}=|H^L_{-\frac{1}{2},0}|^2+|H^L_{-\frac{1}{2},-}|^2+\frac{m_\ell^2}{2q^2}\Big(3|H^L_{-\frac{1}{2},t}|^2+|H^L_{-\frac{1}{2},0}|^2+|H^L_{-\frac{1}{2},-}|^2\Big);\\
     &B^{\lambda_2=-\frac{1}{2}}_{VR}=|H^R_{-\frac{1}{2},0}|^2+|H^R_{-\frac{1}{2},-}|^2+\frac{m_\ell^2}{2q^2}\Big(3|H^R_{-\frac{1}{2},t}|^2+|H^R_{-\frac{1}{2},0}|^2+|H^R_{-\frac{1}{2},-}|^2\Big);\\
     &B^{\lambda_2=-\frac{1}{2}}_{SL}=\frac{3}{2}|H^{SPL}_{-\frac{1}{2}}|^2;\\
     &B^{\lambda_2=-\frac{1}{2}}_{SR}=\frac{3}{2}|H^{SPR}_{-\frac{1}{2}}|^2;\\
     &B^{\lambda_2=-\frac{1}{2}}_{T}=2\bigg[\Big(H^T_{-\frac{1}{2},+,-}-H^T_{-\frac{1}{2},0,t}\Big)^2+\Big(H^T_{-\frac{1}{2},0,-}-H^T_{-\frac{1}{2},-,t}\Big)^2;\nonumber\\
    &\,\,\,\,\,\,\,\,\,\,\,\,\,\,\,\,\,\,\,\,\,\,\,\,\,\,\,\,\,\,
     +\frac{2m_\ell^2}{q^2}\bigg(\Big(H^T_{-\frac{1}{2},+,-}-H^T_{-\frac{1}{2},0,t}\Big)^2+\Big(H^T_{-\frac{1}{2},0,-}-H^T_{-\frac{1}{2},-,t}\Big)^2\bigg)\bigg];\\
     &B^{\lambda_2=-\frac{1}{2}}_{VL,VR}=H^L_{-\frac{1}{2},0}H^R_{-\frac{1}{2},0}+H^L_{-\frac{1}{2},-}H^R_{-\frac{1}{2},-}+\frac{m_\ell^2}{2q^2}\Big(3H^L_{-\frac{1}{2},t}H^R_{-\frac{1}{2},t}+H^L_{-\frac{1}{2},0}H^R_{-\frac{1}{2},0}+H^L_{-\frac{1}{2},-}H^R_{-\frac{1}{2},-}\Big);\\
     &B^{\lambda_2=-\frac{1}{2}}_{VL,SL}=-\frac{3m_\ell}{2\sqrt{q^2}}H^L_{-\frac{1}{2},t}H^{SPL}_{-\frac{1}{2}};\\
     &B^{\lambda_2=-\frac{1}{2}}_{VL,SR}=-\frac{3m_\ell}{2\sqrt{q^2}}H^L_{-\frac{1}{2},t}H^{SPR}_{-\frac{1}{2}};\\
     &B^{\lambda_2=-\frac{1}{2}}_{VL,T}=\frac{3m_\ell}{\sqrt{q^2}}\Big[H^L_{-\frac{1}{2},0}\Big(H^T_{-\frac{1}{2},+,-}-H^T_{-\frac{1}{2},0,t}\Big)+H^L_{-\frac{1}{2},-}\Big(H^T_{-\frac{1}{2},0,-}-H^T_{-\frac{1}{2},-,t}\Big)\Big];\\
     &B^{\lambda_2=-\frac{1}{2}}_{VR,SL}=-\frac{3m_\ell}{2\sqrt{q^2}}H^R_{-\frac{1}{2},t}H^{SPL}_{-\frac{1}{2}};\\
     &B^{\lambda_2=-\frac{1}{2}}_{VR,SR}=-\frac{3m_\ell}{2\sqrt{q^2}}H^R_{-\frac{1}{2},t}H^{SPR}_{-\frac{1}{2}};\\
     &B^{\lambda_2=-\frac{1}{2}}_{VR,T}=\frac{3m_\ell}{\sqrt{q^2}}\Big[H^R_{-\frac{1}{2},0}\Big(H^T_{-\frac{1}{2},+,-}-H^T_{-\frac{1}{2},0,t}\Big)+H^R_{-\frac{1}{2},-}\Big(H^T_{-\frac{1}{2},0,-}-H^T_{-\frac{1}{2},-,t}\Big)\Big];\\
     &B^{\lambda_2=-\frac{1}{2}}_{SL,SR}=\frac{3}{2}H^{SPL}_{-\frac{1}{2}}H^{SPR}_{-\frac{1}{2}};\\
     &B^{\lambda_2=-\frac{1}{2}}_{SL,T}=0;\\
     &B^{\lambda_2=-\frac{1}{2}}_{SR,T}=0.
\end{align}

Similarly, we also obtained the expression of $\frac{d\Gamma^{\lambda_\ell}}{dq^2}$ just by replacing $B$ functions by $F$ functions, which are given as
\begin{align}
    &F^{\lambda_{\ell}=\frac{1}{2}}_{VL}=|H^L_{\frac{1}{2},+}|^2+|H^L_{-\frac{1}{2},0}|^2+|H^L_{\frac{1}{2},0}|^2+|H^L_{-\frac{1}{2},-}|^2;\\
    &F^{\lambda_{\ell}=\frac{1}{2}}_{VR}=|H^R_{\frac{1}{2},+}|^2+|H^R_{-\frac{1}{2},0}|^2+|H^R_{\frac{1}{2},0}|^2+|H^R_{-\frac{1}{2},-}|^2;\\
    &F^{\lambda_{\ell}=\frac{1}{2}}_{SL}=0;\\
    &F^{\lambda_{\ell}=\frac{1}{2}}_{SR}=0;\\
    &F^{\lambda_{\ell}=\frac{1}{2}}_{T}=\frac{4m_\ell^2}{q^2}\Big[\Big(H^T_{\frac{1}{2},+,t}-H^T_{\frac{1}{2},+,0}\Big)^2+\Big(H^T_{-\frac{1}{2},+,-}-H^T_{-\frac{1}{2},0,t}\Big)^2\nonumber\\
    &\,\,\,\,\,\,\,\,\,\,\,\,\,\,\,\,\,\,\,\,\,\,\,\,\,\,\,\,\,\,
    +\Big(H^T_{\frac{1}{2},+,-}-H^T_{\frac{1}{2},0,t}\Big)^2+\Big(H^T_{-\frac{1}{2},0,-}-H^T_{-\frac{1}{2},-,t}\Big)^2\Big];\\
    &F^{\lambda_{\ell}=\frac{1}{2}}_{VL,VR}=H^L_{\frac{1}{2},+}H^R_{\frac{1}{2},+}+H^L_{-\frac{1}{2},0}H^R_{-\frac{1}{2},0}+H^L_{\frac{1}{2},0}H^R_{\frac{1}{2},0}+H^L_{-\frac{1}{2},-}H^R_{-\frac{1}{2},-};\\
    &F^{\lambda_{\ell}=\frac{1}{2}}_{VL,SL}=0;\\
    &F^{\lambda_{\ell}=\frac{1}{2}}_{VL,SR}=0;\\
    &F^{\lambda_{\ell}=\frac{1}{2}}_{VL,T}=-\frac{2m_\ell}{\sqrt{q^2}}\Big[H^L_{\frac{1}{2},+}\Big(H^T_{\frac{1}{2},+,t}-H^T_{\frac{1}{2},+,0}\Big)+H^L_{-\frac{1}{2},0}\Big(H^T_{-\frac{1}{2},0,t}-H^T_{-\frac{1}{2},+,-}\Big)+H^L_{\frac{1}{2},0}\Big(H^T_{\frac{1}{2},0,t}-H^T_{\frac{1}{2},+,-}\Big);\\
    &+H^L_{-\frac{1}{2},-}\Big(H^T_{-\frac{1}{2},-,t}-H^T_{-\frac{1}{2},0,-}\Big)\Big];\\
    &F^{\lambda_{\ell}=\frac{1}{2}}_{VR,SL}=0;\\
    &F^{\lambda_{\ell}=\frac{1}{2}}_{VR,SR}=0;\\
    &F^{\lambda_{\ell}=\frac{1}{2}}_{VR,T}=-\frac{2m_\ell}{\sqrt{q^2}}\Big[H^R_{\frac{1}{2},+}\Big(H^T_{\frac{1}{2},+,t}-H^T_{\frac{1}{2},+,0}\Big)+H^R_{-\frac{1}{2},0}\Big(H^T_{-\frac{1}{2},0,t}-H^T_{-\frac{1}{2},+,-}\Big)+H^R_{\frac{1}{2},0}\Big(H^T_{\frac{1}{2},0,t}-H^T_{\frac{1}{2},+,-}\Big);\\
    &+H^R_{-\frac{1}{2},-}\Big(H^T_{-\frac{1}{2},-,t}-H^T_{-\frac{1}{2},0,-}\Big)\Big];\\
    &F^{\lambda_{\ell}=\frac{1}{2}}_{SL,SR}=0;\\
    &F^{\lambda_{\ell}=\frac{1}{2}}_{SL,T}=0;\\
    &F^{\lambda_{\ell}=\frac{1}{2}}_{SR,T}=0;\\
    &F^{\lambda_{\ell}=-\frac{1}{2}}_{VL}=\frac{m_\ell^2}{2q^2}\Big[3|H^L_{-\frac{1}{2},t}|^2+3|H^L_{\frac{1}{2},t}|^2+|H^L_{\frac{1}{2},+}|^2+|H^L_{-\frac{1}{2},0}|^2+|H^L_{\frac{1}{2},0}|^2+|H^L_{-\frac{1}{2},-}|^2\Big];\\
    &F^{\lambda_{\ell}=-\frac{1}{2}}_{VR}=\frac{m_\ell^2}{2q^2}\Big[3|H^R_{-\frac{1}{2},t}|^2+3|H^R_{\frac{1}{2},t}|^2+|H^R_{\frac{1}{2},+}|^2+|H^R_{-\frac{1}{2},0}|^2+|H^R_{\frac{1}{2},0}|^2+|H^R_{-\frac{1}{2},-}|^2\Big];\\
    &F^{\lambda_{\ell}=-\frac{1}{2}}_{SL}=\frac{3}{2}\Big(|H^{SPL}_{-\frac{1}{2}}|^2+|H^{SPL}_{\frac{1}{2}}|^2\Big);\\
    &F^{\lambda_{\ell}=-\frac{1}{2}}_{SR}=\frac{3}{2}\Big(|H^{SPR}_{-\frac{1}{2}}|^2+|H^{SPR}_{\frac{1}{2}}|^2\Big);\\
    &F^{\lambda_{\ell}=-\frac{1}{2}}_{T}=2\Big[\Big(H^T_{\frac{1}{2},+,t}-H^T_{\frac{1}{2},+,0}\Big)^2+\Big(H^T_{-\frac{1}{2},+,-}-H^T_{-\frac{1}{2},0,t}\Big)^2\nonumber\\
    &\,\,\,\,\,\,\,\,\,\,\,\,\,\,\,\,\,\,\,\,\,\,\,\,\,\,\,\,\,\,+\Big(H^T_{\frac{1}{2},+,-}-H^T_{\frac{1}{2},0,t}\Big)^2+\Big(H^T_{-\frac{1}{2},0,-}-H^T_{-\frac{1}{2},-,t}\Big)^2\Big];\\
    &F^{\lambda_{\ell}=-\frac{1}{2}}_{VL,VR}=\frac{m_\ell^2}{2q^2}\Big[3H^L_{-\frac{1}{2},t}H^R_{-\frac{1}{2},t}+3H^L_{\frac{1}{2},t}H^R_{\frac{1}{2},t}+H^L_{\frac{1}{2},+}H^R_{\frac{1}{2},+}+H^L_{-\frac{1}{2},0}H^R_{-\frac{1}{2},0}+H^L_{\frac{1}{2},0}H^R_{\frac{1}{2},0}+H^L_{-\frac{1}{2},-}H^R_{-\frac{1}{2},-}\Big];\\
    &F^{\lambda_{\ell}=-\frac{1}{2}}_{VL,SL}=-\frac{3m_\ell}{2\sqrt{q^2}}\Big[H^L_{-\frac{1}{2},t}H^{SPL}_{-\frac{1}{2}}+H^L_{\frac{1}{2},t}H^{SPL}_{\frac{1}{2}}\Big];\\
    &F^{\lambda_{\ell}=-\frac{1}{2}}_{VL,SR}=-\frac{3m_\ell}{2\sqrt{q^2}}\Big[H^L_{-\frac{1}{2},t}H^{SPR}_{-\frac{1}{2}}+H^L_{\frac{1}{2},t}H^{SPR}_{\frac{1}{2}}\Big];\\
    &F^{\lambda_{\ell}=-\frac{1}{2}}_{VL,T}=-\frac{m_\ell}{\sqrt{q^2}}\Big[H^L_{\frac{1}{2},+}\Big(H^T_{\frac{1}{2},+,t}-H^T_{\frac{1}{2},+,0}\Big)+H^L_{-\frac{1}{2},0}\Big(H^T_{-\frac{1}{2},0,t}-H^T_{-\frac{1}{2},+,-}\Big)\nonumber\\
    &\,\,\,\,\,\,\,\,\,\,\,\,\,\,\,\,\,\,\,\,\,\,\,\,\,\,\,\,\,\,
    +H^L_{\frac{1}{2},0}\Big(H^T_{\frac{1}{2},0,t}-H^T_{\frac{1}{2},+,-}\Big)+H^L_{-\frac{1}{2},-}\Big(H^T_{-\frac{1}{2},-,t}-H^T_{-\frac{1}{2},0,-}\Big)\Big];\\
    &F^{\lambda_{\ell}=-\frac{1}{2}}_{VR,SL}=-\frac{3m_\ell}{2\sqrt{q^2}}\Big[H^R_{-\frac{1}{2},t}H^{SPL}_{-\frac{1}{2}}+H^R_{\frac{1}{2},t}H^{SPL}_{\frac{1}{2}}\Big];\\
    &F^{\lambda_{\ell}=-\frac{1}{2}}_{VR,SR}=-\frac{3m_\ell}{2\sqrt{q^2}}\Big[H^R_{-\frac{1}{2},t}H^{SPR}_{-\frac{1}{2}}+H^R_{\frac{1}{2},t}H^{SPR}_{\frac{1}{2}}\Big];\\
    &F^{\lambda_{\ell}=-\frac{1}{2}}_{VR,T}=-\frac{m_\ell}{\sqrt{q^2}}\Big[H^R_{\frac{1}{2},+}\Big(H^T_{\frac{1}{2},+,t}-H^T_{\frac{1}{2},+,0}\Big)+H^R_{-\frac{1}{2},0}\Big(H^T_{-\frac{1}{2},0,t}-H^T_{-\frac{1}{2},+,-}\Big)\nonumber\\
    &\,\,\,\,\,\,\,\,\,\,\,\,\,\,\,\,\,\,\,\,\,\,\,\,\,\,\,\,\,\,
    +H^R_{\frac{1}{2},0}\Big(H^T_{\frac{1}{2},0,t}-H^T_{\frac{1}{2},+,-}\Big)+H^R_{-\frac{1}{2},-}\Big(H^T_{-\frac{1}{2},-,t}-H^T_{-\frac{1}{2},0,-}\Big)\Big];\\
    &F^{\lambda_{\ell}=-\frac{1}{2}}_{SL,SR}=\frac{3}{2}\Big[H^{SPL}_{-\frac{1}{2}}H^{SPR}_{-\frac{1}{2}}+H^{SPL}_{\frac{1}{2}}H^{SPR}_{\frac{1}{2}}\Big];\\
    &F^{\lambda_{\ell}=-\frac{1}{2}}_{SL,T}=0;\\
    &F^{\lambda_{\ell}=-\frac{1}{2}}_{SR,T}=0.
\end{align}
\end{appendix}
\bibliographystyle{bibstyle}
\bibliography{mybib}
\end{document}